\documentclass[apj]{emulateapj}

\usepackage{graphicx}
\usepackage{epsfig}

\newcommand{\fesc}{\ifmmode{f_{\rm esc}}\else{$f_{\rm esc}$}\fi}
\newcommand{\fescs}{\ifmmode{f_{\rm esc}^\star}\else{$f_{\rm esc}^\star$}\fi}
\newcommand{\kms}{\ifmmode{~{\rm km~s^{-1}}}\else{~km s$^{-1}$}\fi}
\newcommand{\cubecm}{\ifmmode{~{\rm cm^{-3}}}\else{~cm$^{-3}$}\fi}
\newcommand{\lsim}{\lower0.3em\hbox{$\,\buildrel <\over\sim\,$}}
\newcommand{\gsim}{\lower0.3em\hbox{$\,\buildrel >\over\sim\,$}}

\newcommand{\flux}{erg s$^{-1}$ cm$^{-2}$ Hz$^{-1}$}
\newcommand{\emis}{erg s$^{-1}$ cm$^{-2}$ Hz$^{-1}$ sr$^{-1}$}

\newcommand{\enzo}{{\sl Enzo}}
\newcommand{\Ms}{\ifmmode{M_\odot}\else{$M_\odot$}\fi}

\newcommand{\hh}{H$_2$}
\newcommand{\Ol}{$\Omega_\Lambda$}
\newcommand{\Om}{$\Omega_M$}
\newcommand{\Ob}{$\Omega_b$}

\newcommand{\tcool}{$t_{\rm{cool}}$}

\newcommand{\tdyn}{$t_{\rm{dyn}}$}

\newcommand{\tvir}{\ifmmode{T_{\rm{vir}}}\else{$T_{\rm{vir}}$}\fi}
\newcommand{\mvir}{\ifmmode{M_{\rm{vir}}}\else{$M_{\rm{vir}}$}\fi}
\newcommand{\rvir}{\ifmmode{r_{\rm{vir}}}\else{$r_{\rm{vir}}$}\fi}

\newcommand{\lya}{Ly$\alpha$}
\newcommand{\jj}{\ifmmode{J_{21}}\else{$J_{21}$}\fi}
\newcommand{\flw}{\ifmmode{F_{LW}}\else{$F_{LW}$}\fi}

\newcommand{\msun}{{\rm\,M_\odot}}

\newcommand\tento[1]{$10^{#1}$}

\begin{document}

\shorttitle{PHOTON ESCAPE FRACTIONS FROM EARLY DWARF GALAXIES}
\shortauthors{WISE \& CEN}

\title{Ionizing Photon Escape Fractions from High Redshift Dwarf Galaxies}
\author{John H. Wise\altaffilmark{1,2} and Renyue Cen\altaffilmark{3}}

\altaffiltext{1}{Laboratory for Observational Cosmology, NASA Goddard
  Space Flight Center, Greenbelt, MD 21114} 
\altaffiltext{2}{NPP Fellow}
\altaffiltext{3}{Department of Astrophysical Sciences, Princeton
  University, Peyton Hall, Ivy Lane, Princeton, NJ 08544}
\email{john.h.wise@nasa.gov}

\begin{abstract}

  It has been argued that low-luminosity dwarf galaxies are the
  dominant source of ionizing radiation during cosmological
  reionization.  The fraction of ionizing radiation that escapes into
  the intergalactic medium from dwarf galaxies with masses less than
  $\sim$$10^{9.5}$ solar masses plays a critical role during this
  epoch.  Using an extensive suite of very high resolution (0.1 pc),
  adaptive mesh refinement, radiation hydrodynamical simulations of
  idealized and cosmological dwarf galaxies, we characterize the
  behavior of the escape fraction in galaxies between $3 \times 10^6$
  and $3 \times 10^9$ solar masses with different spin parameters,
  amounts of turbulence, and baryon mass fractions.  For a given halo
  mass, escape fractions can vary up to a factor of two, depending on
  the initial setup of the idealized halo.  In a cosmological setting,
  we find that the time-averaged photon escape fraction always exceeds
  25\% and reaches up to 80\% in halos with masses above $10^8$ solar
  masses with a top-heavy IMF.  The instantaneous escape fraction can
  vary up to an order of magnitude in a few million years and tend to
  be positively correlated with star formation rate.  We find that the
  mean of the star formation efficiency times ionizing photon escape
  fraction, averaged over all atomic cooling ($T_{\rm vir} \ge
  8000~$K) galaxies, ranges from 0.02 for a normal IMF to 0.03 for a
  top-heavy IMF, whereas smaller, molecular cooling galaxies in
  minihalos do not make a significant contribution to reionizing the
  universe due to a much lower star formation efficiency.  These
  results provide the physical basis for cosmological reionization by
  stellar sources, predominately atomic cooling dwarf galaxies.
  
\end{abstract}


\section{Introduction}

In most calculations of cosmological reionization, it has to be
assumed that the product of star formation efficiency and hydrogen
ionizing photon escape fraction is $\ge 0.01$ \citep[e.g.][]{Gnedin00,
  Cen03a, Cen03b, Wyithe03a, Wyithe03b, Venkatesan03, Somerville03,
  Chiu03, Haiman03, Ciardi03, Sokasian03, Sokasian04, Wyithe07,
  Srbinovsky08} in order to reionize the universe early enough to be
consistent with the Wilkinson Microwave Anisotropy Probe
(\textit{WMAP}) observations \citep{Spergel07, Komatsu08}, if stars
produce the majority of ionizing photons.  \citet{Gnedin00} suggested
that, based on Local Group dwarf galaxies, the star formation
efficiency at high redshift is $\sim 4\%$, consistent with theoretical
works \citep[e.g.][]{Krumholz05, Krumholz07}.  Thus, unless star
formation efficiency is unusually high (i.e., $\ge 10\%$) at high
redshift, this requirement seems to suggest a high ionizing photon
escape fraction from high redshift galaxies, $f_{\rm esc} \ge 10\%$,
may be necessary in the context of stellar reionization within the
standard cold dark matter model.

However, $f_{\rm esc} \ge 10\%$ is by no means the norm, at least from
observations at lower redshifts.  Approximately 6\% of ionizing
radiation escape from the Milky Way \citep{Bland99, Putman03}.  For
local starburst galaxies \citet{Hurwitz97} gave $f_{\rm esc}\le
0.032,0.052, 0.11$ ($2\sigma$) for Mrk 496, Mrk 1267, and IRAS
08339+6517 ($\le 0.57$ in the case of Mrk 66).  \citet{Deharveng01}
gave an escape fraction of $f_{\rm esc}<0.062$ for Mrk 54.
\citet{Heckman01} found $f_{\rm esc} \le 0.06$.  \citet{Bergvall06}
found $f_{\rm esc}\sim 0.04-0.1$ for a local extreme starburst dwarf,
the Blue Compact Galaxy Haro 11.  \citet{Chen07} placed a 95\% upper
limit of $f_{\rm esc}\le 0.075$ for star forming regions hosting
gamma-ray bursts at $z\ge 2$.  For Lyman break galaxies at $z\sim 3$,
\citet{Shapley06} found $f_{\rm esc}\sim 0.03$.  \citet{Siana07}
examined starburst galaxies at $z \sim 1.3$ and found that less than
20\% have relative escape fractions%
\footnote{This quantity is the absolute escape fraction that is
  corrected for dust attenuation at 1500 \AA~and is a common measure
  in observations of escaping Lyman continuum.} near unity and a
global absolute escape fraction of $\lsim 0.04$.  However in this
study, some galaxies have upper limits of $f_{\rm esc, rel}$ as low as
0.08.  \citet{Inoue06} recently compiled available observations over a
wide range of redshift and come to the conclusion that there is a
trend of increasing $f_{\rm esc}$ with redshift from $\le 0.01$ at
$z\le 1$ to $\sim 0.1$ at $z\ge 4$.

Some theoretical works also seem to point to low values for $f_{\rm
  esc}$.  Theoretical models of our own Galaxy with realistic star
formation mode by \citet{Dove00} give an estimate of $f_{\rm esc}\sim
0.06$ or lower.  \citet{Ciardi02} studied the effect of gas
inhomogeneities on $f_{\rm esc}$ for Milky Way like galaxies using
simulations with radiative transfer and concluded that $f_{\rm esc}$
depends on the density structure as well as star formation rate (SFR).
In addition, porosity in the interstellar medium (ISM) that is caused
by SN mechanical feedback may provide additional channels in which
radiation could escape, thus increasing the UV escape fraction
\citep{Clarke02}.  \citet{Wood00} argued that $f_{\rm esc}\le 0.01$
for galaxies at $z\sim 10$.  \citet{Ricotti00} found that $f_{\rm
  esc}\ge 0.1$ only for halos of total mass less than $\sim 10^7\msun$
at $z\ge 6$ with $f_{\rm esc}$ dropping precipitously for larger
halos.  \citet{Fujita03}, using ZEUS-3D simulations, found that
$f_{\rm esc}\le 0.1$ from disks of dwarf starburst galaxies of total
mass $M\ge 10^{8-10}\msun$.  \citet[][hereafter RS06 and
RS07]{Razoumov06, Razoumov07} concluded that $f_{\rm esc}$ = 0.01--0.1
in several young galaxies with $\mvir = 10^{12-13} \Ms$ at $z \sim 3$
in their simulations with radiative transfer using adaptive ray
tracing, agreeing with the observations presented in \citet{Inoue06}.
\citet[][hereafter GKC08]{Gnedin08}, using detailed hydrodynamic
simulations with radiative transfer, found $f_{\rm esc}\sim 0.01-0.03$
for galaxies of total mass $M\ge 10^{11}\msun$ at $z=3-5$.

In the relevant redshift range for cosmological reionization,
$z=6-15$, most of the ionizing radiation due to stars comes from dwarf
galaxies of $M\le 10^{8-9}\msun$ in the standard cold dark matter
model \citep{Barkana01}.  This purpose of this paper is to investigate
how $f_{\rm esc}$ depends on various physical parameters expected in
realistic cosmological settings using detailed adaptive mesh
refinement (AMR) simulations coupled with adaptive 3D ray-tracing
radiative transfer, focusing on galaxies with total mass
$M=10^{6.5}-10^{9.5}\msun$ at $z\ge 6$.  We study the dependence of
$f_{\rm esc}$ on four physical parameters: mass of the galaxy, spin
parameter, baryonic mass fraction and turbulent energy.  Our work is
an extension of GKC08 and \citet{Fujita03}, by exploring the
dependence of $f_{\rm esc}$ on some important physical parameters
aforementioned, by including lower mass galaxies and by employing very
high resolution ($0.1$pc) simulations.  In addition, importantly, we
allow for disparate lifetimes of stars of different masses, of their
subsequent explosive energy (for those stars that become supernovae)
input into the ISM.

We first describe our radiation hydrodynamics simulations of isolated
and cosmological halos in \S\ref{sec:sims}.  There we also describe
our algorithm for star formation and feedback that considers a
multi-phase ISM and resolved molecular clouds.  In \S\ref{sec:sf} and
\S\ref{sec:fesc}, we present the star formation rates and history and
the resulting escape fraction of UV radiation, respectively.  Next in
\S\ref{sec:disc}, we further compare our results with previous work
and discuss effects from any neglected physical processes in our
simulations and the implications of our results on reionization
scenarios.  We summarize our work in the last section.


\section{Radiation Hydrodynamical Simulations}
\label{sec:sims}

We use the Eulerian adaptive mesh refinement (AMR) hydrodynamic code
\enzo~\citep{Bryan97, Bryan99, OShea04} to investigate the escape of
ionizing radiation from early dwarf galaxies.  We have modified the
code to include a time-dependent implementation of radiative transfer
that adaptively ray traces from point sources \citep{Abel02b, Abel07,
  Wise08a}.  \enzo~uses an $n$-body adaptive particle-mesh solver
\citep{Efstathiou85} to follow the dark matter (DM) dynamics.  It
solves the hydrodynamical equation using the second-order accurate
piecewise parabolic method \citep{Woodward84, Bryan95}, while a
Riemann solver ensures accurate shock capturing with minimal
viscosity.  We use the nine-species (\ion{H}{1}, \ion{H}{2},
\ion{He}{1}, \ion{He}{2}, \ion{He}{3}, e$^-$, \hh, \hh$^+$, H$^-$)
non-equilibrium chemistry model in \enzo~\citep{Abel97, Anninos97} and
the \hh~cooling rates from \citet{Galli98}.  We adopt the cosmological
parameters from the third year ``mean'' WMAP results \citep{Spergel07}
of \Om~= 1~--~\Ol~= 0.24, \Ob$h^2$ = 0.0229, and $h$ = 0.73, where
\Ol, \Om, and \Ob~are the fractions of mass-energy contained in vacuum
energy, cold dark matter, and baryons, respectively.  Here $h$ is the
Hubble constant in units of 100 km~s$^{-1}$ Mpc$^{-1}$.  For the
cosmological simulations, we use $\sigma_8 = 0.76$ and $n$ = 0.96,
where $\sigma_8$ is the $rms$ of density fluctuations inside a sphere
of radius 8$h^{-1}$ Mpc and $n$ is the scalar spectral index.

We first describe our simulation setup of our suite of idealized dwarf
galaxies.  Next we detail our cosmological simulations that we use to
check the validity of our assumptions in the idealized cases.  Last we
explain our star formation and feedback model.

%
%

\begin{deluxetable*}{ccccccccccccc}
\tabletypesize{\footnotesize}
\tablewidth{0pc}
\tablecaption{Cosmological Halo Properties\label{tab:halos}}

\tablehead{
  \colhead{} & 
  \colhead{} & 
  \colhead{} &
  \colhead{} &
  \colhead{} &
  \colhead{} &
  \colhead{} &
  \multicolumn{3}{c}{High Luminosity} &
  \multicolumn{3}{c}{Normal Luminosity} \\
\cline{8-10} \cline{11-13} \\
  \colhead{\#} & 
  \colhead{\mvir} & 
  \colhead{$f_b$} &
  \colhead{$f_{turb}$} &
  \colhead{$\lambda$} &
  \colhead{$N_{\rm{grid,0}}$} &
  \colhead{$N_{\rm{part}}$} &
  \colhead{$\langle f_{\rm{esc}} \rangle$} &
  \colhead{$\langle$SFR$\rangle$} &
  \colhead{max(SFR)} &
  \colhead{$\langle f_{\rm{esc}} \rangle$} &
  \colhead{$\langle$SFR$\rangle$} &
  \colhead{max(SFR)} \\
  \colhead{} & 
  \colhead{[\Ms]} & 
  \colhead{} & 
  \colhead{} & 
  \colhead{} & 
  \colhead{} &
  \colhead{} & 
  \colhead{} &
  \colhead{[\Ms\ yr$^{-1}$]} & 
  \colhead{[\Ms\ yr$^{-1}$]} &
  \colhead{} &
  \colhead{[\Ms\ yr$^{-1}$]} & 
  \colhead{[\Ms\ yr$^{-1}$]} \\
  \colhead{(1)} & 
  \colhead{(2)} & 
  \colhead{(3)} & 
  \colhead{(4)} & 
  \colhead{(5)} & 
  \colhead{(6)} &
  \colhead{(7)} & 
  \colhead{(8)} &
  \colhead{(9)} & 
  \colhead{(10)} &
  \colhead{(11)} &
  \colhead{(12)} & 
  \colhead{(13)}
}
\startdata
1 & $1.6 \times 10^7$ & 0.11 & 0.56 & 0.057 & 118$^3$ & 5,144 & 0.38 & $3.7 \times
10^{-5}$ & $1.7 \times 10^{-4}$ & 0.065 & $1.5 \times 10^{-4}$ & $4.1
\times 10^{-3}$ \\
2 & $2.6 \times 10^7$ & 0.12 & 0.48 & 0.038 & 130$^3$ & 8,482 & 0.77 & $3.0 \times
10^{-4}$ & $1.1 \times 10^{-3}$ & 0.13 & $6.7 \times 10^{-4}$ & $4.8
\times 10^{-3}$ \\
3 & $2.7 \times 10^7$ & 0.13 & 0.61 & 0.061 & 132$^3$ & 8,809 & 0.42 & $1.0 \times
10^{-3}$ & $3.7 \times 10^{-3}$ & 0.11 & $9.5 \times 10^{-4}$ & $3.1
\times 10^{-3}$ \\
4 & $5.0 \times 10^7$ & 0.13 & 0.57 & 0.013 & 154$^3$ & 16,103 & 0.67 & $1.1 \times
10^{-3}$ & $4.1 \times 10^{-3}$ & 0.44 & $4.1 \times 10^{-3}$ & $1.6
\times 10^{-2}$ \\
5 & $1.7 \times 10^8$ & 0.14 & 0.67 & 0.044 & 240$^3$ & 52,831 & 0.78 & $2.6 \times
10^{-2}$ & $7.6 \times 10^{-2}$ & 0.38 & $1.6 \times 10^{-2}$ & $6.5
\times 10^{-2}$ \\
6 & $3.8 \times 10^8$ & 0.12 & 0.60 & 0.027 & 80$^3$ & 1,222 & 0.44 & $2.3 \times
10^{-2}$ & $8.9 \times 10^{-2}$ & 0.39 & $5.0 \times 10^{-2}$ & $1.1
\times 10^{-1}$ \\
7 & $6.7 \times 10^8$ & 0.12 & 0.58 & 0.052 & 84$^3$ & 2,184 & 0.72 & $5.3 \times
10^{-2}$ & $2.2 \times 10^{-1}$ & 0.36 & $5.4 \times 10^{-2}$ & $1.7
\times 10^{-1}$ \\
8 & $8.7 \times 10^8$ & 0.13 & 0.54 & 0.025 & 90$^3$ & 2,783 & 0.47 & $4.8 \times
10^{-2}$ & $1.9 \times 10^{-1}$ & 0.31 & $6.3 \times 10^{-2}$ & $2.2
\times 10^{-1}$ \\
9 & $1.2 \times 10^9$ & 0.13 & 0.62 & 0.049 & 98$^3$ & 3,863 & 0.51 & $5.7 \times
10^{-2}$ & $1.9 \times 10^{-1}$ & 0.40 & $5.3 \times 10^{-2}$ & $1.1
\times 10^{-1}$ \\
10& $4.0 \times 10^9$ & 0.13 & 0.70 & 0.069 & 156$^3$ & 12,796 & 0.62 & $1.7 \times
10^{-1}$ & $5.1 \times 10^{-1}$ & 0.54 & $2.2 \times 10^{-1}$ & $4.2
\times 10^{-1}$
\enddata

\tablecomments{Low and high luminosity models have 2,600 and 26,000
  ionizing photons per stellar baryon.  Col. (1): Halo
  number. Col. (2): Virial mass. Col. (3): Baryon mass fraction.
  Col. (4): Turbulent energy.  Col. (5): Spin parameter.  Col. (6):
  Number of grid cells on the top-level AMR grid.  Col. (7): Number of
  dark matter particles inside the virial radius. Col. (8,11):
  Time-averaged UV escape fraction. Col. (9,12): Time-averaged star
  formation rate.  Col. (10,13): Maximum star formation rate.}
\end{deluxetable*}

\subsection{Isolated Halo Setup}

In order to quantify the dependence of \fesc~on dwarf galaxy
properties, we run a total of 75 idealized dwarf galaxy simulations,
in which we vary the total halo mass \mvir, spin parameter
\begin{equation}
  \label{eq:spin}
  \lambda = \frac{L \vert E \vert^{1/2}}{GM_{\rm{vir}}^{5/2}},
\end{equation}
baryon mass fraction $f_b = M_{\rm{gas}} / \mvir$, and gas turbulence
$f_{\rm{turb}} = v_{rms} / V_c$.  Here $L$, $E$, and $M_{\rm{gas}}$
are the total angular momentum, energy, and gas mass in the halo.
When initializing the halos, $E$ is the total gravitational energy of
the halo, and during the analysis, $E = M v_{\rm rms}^2 / 2$.
$v_{rms}$ is the three-dimensional $rms$ gas velocity, and
\begin{equation}
  \label{eq:vc}
  V_c = \left(\frac{GM_{\rm{vir}}}{r_{\rm{vir}}}\right)^{1/2}
\end{equation}
is the circular velocity of the halo, where \rvir~is the virial
radius.  
We define the virial radius and thus mass as the radius of a
sphere that encloses an overdensity of 200 times the cosmic mean
density.  

We calibrate our choice of such parameters by the following previous
work.
\begin{enumerate}
\item \textit{Spin parameter}--- It has been well-established
  \citep{Barnes87, Eisenstein95} that the distribution of the halo
  spin parameter is log-normal with a mean of 0.042 and deviation of
  0.5 \citep[e.g.][]{Bullock01a}.
\item \textit{Baryon fraction}--- Radiative feedback from the first
  stars and subsequent episodes of star formation generate outflows
  that reduce the baryon fraction up to a factor of 3 below the cosmic
  average (\Ob/\Om) in halos with $\mvir \sim 10^8 \Ms$
  \citep{Ricotti05, Wise08a, Ricotti08, Mesigner08}.
\item \textit{Turbulent energy}--- Turbulence increases during the
  assembly of a cosmological halo through both virialization and
  mergers.  Typical Mach numbers in halos with an adiabatic state of
  equation is $\sim$0.25 \citep{Norman99, Nagai03, Wise07a}.  In halos
  with gas that can efficiently cool, the halo cannot no longer
  virialize through gas heating, becoming more turbulent to virialize.
  The combination of lower temperatures and greater turbulent motions
  leads to turbulent Mach numbers above unity and close to the
  circular velocity of the halo, i.e. $f_{\rm{turb}} = 1$
  \citep{Wise07a, Greif08}.
\end{enumerate}
We hence fix two of the following parameters ($\lambda$, $f_b$,
$f_{\rm{turb}}$) = (0.04, 0.1, 0.25) and change the remaining parameter
and halo mass in each calculation.  We allow these halo parameters to
vary as:
\begin{itemize}
\item log$_{10}$ \mvir~= 6.5, 7, 7.5, 8, 8.5, 9, 9.5;
\item $\lambda$ = 0, 0.02, 0.04, 0.06, 0.1;
\item $f_b$ = 0.05, 0.075, 0.1, 0.15;
\item $f_{\rm{turb}}$ = 0, 0.25, 0.5, 1.
\end{itemize}
For example, when varying $f_b$, we model halos with all of the quoted
virial masses and $f_b$ while fixing $\lambda$ and $f_{\rm{turb}}$ to
0.04 and 0.25, respectively, resulting in a total of 28 calculations.
We repeat this method for varying the spin parameter and turbulent
energy.  We did not evaluate a $\mvir = 10^{6.5} \Ms$ halo in two cases,
$f_b = 0.05$ and $f_{turb} = 1$, because they did not condense
to form stars.

The simulation box has a side length of 10 times the virial radius at
a virialization redshift $z_{vir} = 8$ and a top grid resolution of
128$^3$.  We refine the grid structure if the gas density becomes 1.5
times greater than the mean gas density times a factor of $2^l$, where
$l$ is the AMR refinement level.  We also refine to resolve the local
Jeans length by at least 4 cells.  Cells are refined to a maximum AMR
level (ranging from 9--12) that results in a spatial resolution of
$\sim$0.1 proper parsec, which is required to model the formation of
the D-type front at small scales correctly \citep{Whalen04,
  Kitayama04}.  Here we sample the initial Str{\"o}mgren sphere by
several cells across, which has a radius of $\sim$1 pc if we assume an
ionizing luminosity of $10^{50}$ photons s$^{-1}$ and a density of
1800 \cubecm~(see \S\ref{sec:sfalgo}).

We perform the calculations in comoving coordinates.  The
gravitational potential is computed from an NFW density profile, 
\begin{equation}
  \label{eqn:nfw}
  \rho_{\rm{NFW}}(r) = \frac{\rho_s}{(r/r_s)(1+r/r_s)^2} ,
\end{equation}
where $r_s$ is a characteristic inner radius, and $\rho_s$ is the
corresponding inner density this is chosen so the total DM mass within
$r_{\rm{vir}}$ is $(1-f_b)M_{\rm{vir}}$.  The core radius can be
expressed in terms of the concentration parameter $c =
r_s/r_{\rm{vir}}$.  We set $c = 6$, which is in concordance with the
fit of \citet{Bullock01b}.  In addition, we consider self-gravity from
the time-dependent baryon density field.

The grid boundaries are periodic, which has no effect on the evolution
of the galaxy because our simulation box is much larger than the
virial radius.  The initial configuration of each halo is an
isothermal sphere, i.e. $\rho \propto r^{-2}$, with a constant density
core with radius $r_s$.  The initial temperature is set to the virial
temperature.  Rotational energy is modelled as a solid body rotator,
and turbulent energy is initialized as random velocities sampled from
a Maxwellian distribution with a temperature $T = (f_{\rm{turb}}V_c)^2
\mu m_p / k$, where $\mu$ is the mean molecular weight in units of a
proton mass $m_p$.  In most of our models, turbulent energy is
dominant over rotational energy, as shown in \citet{Wise07a}.  We
start our calculations at redshift 8.  The surface of the halo is in
pressure equilibrium with the IGM that has a density of \Ob$\rho_c$,
where $\rho_c = 3H_0^2/8\pi G$ is the critical density of the
universe.

\subsection{Cosmological Halo Setup}

In reality, these halos reside in a cosmological environment, where
adjacent filamentary structures may inhibit the escape of ionizing
radiation.  To investigate any effects of a cosmological setting, we
run two sets of AMR cosmological simulations with periodic boundary
conditions and 384$^3$ particles and grid cells.  The first set of
simulations (named M6.5-8) is set up to study halos with masses
between 10$^{6.5}$ and 10$^8 \Ms$ with a side length of 1.75 comoving
Mpc, resulting in a DM particle mass of 2720 \Ms.  The second set of
simulations (named M8.5-9.5) has a comoving side length of 8 Mpc and
focuses on halos with masses between 10$^{8.5}$ and 10$^{9.5} \Ms$,
resulting in a DM particle mass of $2.72 \times 10^5 \Ms$.  Our choice
of resolution and box size allows halos in the quoted mass range to be
resolved by at least 1000 DM particles.  We initialize simulations
M6.5-8 and M8.5-9.5 at a redshift 68 and 51, respectively, with
\texttt{grafic} \citep{Bertschinger01} and different random phases.
AMR refinement criteria are based on the same quantities as the
idealized halos, now also include DM density, but with an overdensity
criterion of 4 instead of 1.5 for both gas and DM.  We allow the grid
structure to be refined in the entire volume.  These simulations are
stopped at redshift 8, which is the same redshift the idealized halos
are initialized.  Note that GKC08 found no redshift dependence in
\fesc.  At the final redshift, simulations M6.5-8 and M8.5-9.5 have
132,456 and 114,602 AMR grids and $5.63 \times 10^8$ (826$^3$) and
$3.11 \times 10^8$ (677$^3$) unique computational cells, respectively.

We use an adiabatic equation of state in these cosmological
simulations because the inclusion of atomic line and molecular
hydrogen cooling would create star forming, cold, dense molecular
clouds.  Currently it is not feasible to treat $\gsim$300 point
sources with our ray tracing scheme; therefore we suppress any gas
cooling and thus star formation in these large-scale simulations and
take the following approach.

At the final redshift for 10 selected halos with at least 1000 DM
particles, we extract a sub-volume with its AMR hierarchy intact and a
box length of 10 times its virial radius.  The halo masses range from
$1.6 \times 10^7$ to $4.0 \times 10^9 \Ms$, and they do not undergo a
major merger in the 100 Myr after $z = 8$.  We take the resolution of
a level 3 AMR grid in the large-scale simulation to be the top-level
resolution of the sub-volume, ranging from 80$^3$ to 240$^3$.
Additional details of these halos are listed in Table \ref{tab:halos}.
Now each sub-volume is its own simulation, and we consider the
nine-species non-equilibrium chemistry model and star formation and
feedback.  We only refine the grid structure within the inner quarter
of the box that is centered on the halo of interest.  We calculate the
gravitational potential with isolated boundary conditions, and the
potential is assumed to be zero at the boundaries.  We use inflowing
and periodic boundary conditions for baryons and DM particles,
respectively.  We restrict star formation to within 1.5 virial radii
of the halo center.  We adjust the baryon and DM velocities so that
the mass-averaged velocity of the simulation is zero.  We also set the
fractional abundances of (\ion{H}{2}, \ion{He}{2}, \ion{He}{3}, H$^-$,
H$_2$, H$_2^+$) = ($1.2 \times 10^{-5}$, \tento{-14}, \tento{-17}, $2
\times 10^{-9}$, $2 \times 10^{-6}$, $3 \times 10^{-14}$).  Although
these values are not the equilibrium values, the non-equilibrium
chemistry solver quickly converges to the correct values during the
first few timesteps of the calculation.

These halos may represent a lower limit of \fesc~because previous
episodes of star formation may have decreased the baryon mass fraction
and the column densities of adjacent filaments; thus, \fesc~may be
larger in the realistic case that examines both Population II and III
star formation during the entire halo assembly history.  However, we
can use the results from our idealized halos to understand how
\fesc~behaves with different galactic properties and make the
according adjustments to our estimates of \fesc~in these extracted
cosmological halos.



\subsection{Star Formation and Feedback}
\label{sec:sfalgo}

We model star formation through a modified version of the extensively
used algorithm of \citet{Cen92}, where a stellar particle is created
when all of the following criteria are met: (i) an overdensity of
10$^7$, equivalent to 1800$\mu^{-1}$\cubecm~at redshift 8; (ii) a
converging gas velocity field ($\nabla \cdot v < 0$); (iii) rapidly
cooling gas, i.e. the cooling time \tcool~is less than the dynamical
time \tdyn~= 1/$(G\rho)^{1/2}$.  Since we ensure that the local Jeans
length is always sampled by at least 4 cells, a single cell will never
be Jeans unstable\footnote{Refinement based on Jeans length usually
  occurs in moderate refinement levels (i.e. 5--8).  With primordial
  gas cooling, dense gas does not cool below 200 K, and at a density
  of 1800 \cubecm, the corresponding Jeans length is $\sim$5 pc, which
  is well above our resolution limit.}, thus we do not consider the
Jeans unstable criterion outlined in the original work.  When a cell
satisfies the above criteria, we define the star forming cloud as a
sphere with \tdyn~= 3 Myr ($\bar{\rho}_{cl} = 1000\mu \cubecm$) and
radius $R_{cl}$, where a fraction $c_\star = 0.07f_{\rm{cold}}$ of the
gas is gradually converted into a stellar particle, whose final mass
$m_\star = c_\star (4\pi/3) \bar{\rho}_{cl} R_{cl}^3$.  Here
$f_{\rm{cold}}$ is the fraction of gas with $T < 10^4$ K within the
sphere.  This treatment of cold gas accretion is similar to the
multi-phase model of star formation specified in \citet{Springel03}.
We take this approach because we resolve the molecular cloud instead
of employing a subgrid model.  The stellar mass in the particle
increases as
\begin{equation}
  \label{eqn:dMstar}
  \frac{dM}{dt} =
  \left(\frac{m_\star}{t_{\rm{dyn}}}\right)
  \left(\frac{t-t_i}{t_{\rm{dyn}}}\right)
  \exp\left[\frac{-(t-t_i)}{t_{\rm{dyn}}}\right]
\end{equation}
or equivalently
\begin{equation}
  \label{eqn:Mstar}
  \frac{M(t)}{m_\star} = 1 - \frac{(t-t_i) + t_{\rm{dyn}}}{t_{\rm{dyn}}}
  \exp\left[\frac{-(t-t_i)}{t_{\rm{dyn}}}\right].
\end{equation}
We then replace the sphere with a uniform density $\rho_{cl} = (1 -
c_\star)/(G t_{\rm{dyn}}^2)$.

Our choice of the star formation efficiency is in agreement with
\citet{Krumholz05}, who find that this fraction in a free-fall time is
0.077$\alpha_{vir}^{-0.68} \mathcal{M}^{-0.32}$, where $\alpha_{vir}$
is a virialization parameter and is near unity, and $\mathcal{M}$ is
the turbulent Mach number.  We justify our choice of a threshold
\tdyn~with inferred dynamical times from observations of star forming
regions are approximately 700 kyr, and star formation occurs for
several dynamical times \citep[e.g.][]{Tan06}.

In order to minimize the number of radiative sources in the
calculations, we enforce a minimum star particle mass of
$10^{-5}M_{\rm{gas}}$, where $M_{\rm{gas}}$ is the gas mass inside the
halo, and merge stellar particles if they are within $R_{\rm{merge}}$
pc.  To keep the number of sources below $\sim$100, we have found the
relation $R_{\rm{merge}} = 10 (\mvir / 10^8 \Ms)^{0.5}$ pc works well.
The merged particle inherits the center of mass as its position,
mass-averaged velocity, and total mass of its two progenitors.  We
found that the UV escape fraction is not significantly dependent on
this parameter and only varies by $\Delta(f_{\rm esc})/f_{\rm esc}
\sim 0.1$ when we change $R_{\rm{merge}}$ by a factor of 0.1, 0.5, and
2.

Radiative stellar feedback is modelled with an adaptive ray tracing
scheme \citep{Abel02b} that is coupled to the hydrodynamics, energy,
and chemistry solver of Enzo \citep[][Wise et al., in
preparation]{Wise08a}.  It is parallelized with MPI and runs on
distributed and shared memory machines.  Each stellar particle is a
radiative point source and has a constant luminosity of $N_\gamma$ =
26,000 ionizing photons per stellar baryon over its lifetime.  This
ionizing luminosity is appropriate for a Salpeter initial mass
function (IMF) with a metallicity of $4 \times 10^{-4} Z_\odot$ and
lower and upper mass cutoffs of 1 and 150 \Ms~\citep{Schaerer03}.
Similarly high specific luminosities may also be applicable in gas
with $Z \gsim 10^{-2.5} Z_\odot$ at high-redshift whose cooling is
limited to the CMB temperature.  This raises the Jeans mass of such
molecular clouds and may result in a top-heavy IMF with a typical
stellar mass of tens of solar masses \citep{Smith08}.  To compare our
work to \citet{Fujita03} and GKC08 in the ``control halos'' of the
idealized setup and cosmological halos, we use $N_\gamma$ = 2,600 that
is suitable for a solar-metallicity Salpeter-like IMF.  The photons
are evenly distributed among 192 rays, i.e. HEALPix level 2, emanating
from its associated radiation source.  As the rays propagate from the
source, the solid angle that is sampled by each ray increases as
$r^2$.  When this solid angle becomes larger than 20\% of the cell
area that it is transversing, the ray splits according to the
formalism of \citeauthor{Abel02b}.  The rays cast in these simulations
have an energy of 21.5 eV, which is the mean ionizing photon energy
from our fiducial IMF \citep{Schaerer03}.  We have experimented with
photon means energy between 14.0 and 30.0 eV and found that our
results are not sensitive to this energy.  Averaged over 10 Myr, this
results in an ionizing luminosity of 10$^{46}$ and 10$^{47}$ erg
s$^{-1}$ \Ms$^{-1}$ for our chosen stellar luminosities.  We consider
only hydrogen photo-ionization in these simulations.  We do not
consider dust absorption as GKC08 showed it is only important redward
of the Lyman break.  We model the \hh~dissociating radiation between
11.2 and 13.6 eV with an optically-thin $1/r^2$ radiation field
centered on each source with luminosities from
\citeauthor{Schaerer03}.  We use the \hh~photo-dissociation rate
coefficient $k_{diss} = 1.1 \times 10^8 F_{\rm{LW}}\;\rm{s}^{-1}$,
where $F_{\rm{LW}}$ is the \hh~flux in units of \flux~\citep{Abel97}.

After the stellar particle has lived for 4 Myr, we model supernova
(SN) feedback by injecting $(dM/dt)(\Delta t)\; 6.8 \times 10^{48}$
erg $M_\odot^{-1}$ of thermal energy at every timestep of the finest
AMR level into a sphere with a diameter of 1 pc.  If the grid cell
that the star particle resides in is larger than 1 pc, we inject the
energy into the surrounding 27 grid cells for numerical stability.
For SN feedback, we evaluate $dM/dt$ at time, t -- 4 Myr, to model the
SN from the stars that were ``born'' 4 Myr prior to the current time.
We eliminate the star particle when it has an age of 10 Myr,
corresponding to the lifetime of OB-stars that produce the vast
majority of ionizing photons.

\subsection{UV Escape Fraction}

We measure the fraction of escaping UV radiation by comparing the
number of photons that travel beyond the virial radius to the current
stellar luminosity.  In other words, the integral form of the UV
escape fraction is
\begin{equation}
  \label{eqn:fesc}
  f_{\rm esc}(t) \equiv \frac{1}{L_{\rm{UV}}(t)} \int_{r = r_{vir}}
  F_{\rm{UV}}(t) d\mathcal{S},
\end{equation}
where $L_{\rm{UV}}$ and $F_{\rm{UV}}$ are the UV stellar luminosity
and UV flux, respectively, and $\mathcal{S}$ is the spherical surface
at $r = r_{vir}$.  The equivalent quantity that is calculated in the
simulations is
\begin{equation}
  \label{eqn:fesc_dist}
  f_{\rm esc}(t) \equiv \frac{\sum N_i(r_{vir} < r < r_{vir} + c\; dt_p)}
  {\sum N_i(r < c\; dt_p)},
\end{equation}
where $N_i$ is the number of photons contained in the $i$-th photon
package, $r$ is the total distance that the package has travelled, and
$c\:dt_p$ is the distance travelled by a photon within a timestep
$dt_p$.  The numerator sums the number of photons that have travelled
farther than a virial radius within the last timestep, and the
denominator is the number of photons emitted in the last timestep,
which can be easily counted when the photon packages are generated
from the point source.  It is possible to compare these two quantities
that are temporally separated by the light-travel time ($t \sim 3$
kyr), which is much smaller than the dynamical times of the star
formation regions, to which the change in SFR is directly related (see
Eq. \ref{eqn:dMstar}--\ref{eqn:Mstar}).  Also notice that a photon at
a distance $r_{vir}$ from its source does not necessarily correspond
to the surface $\mathcal{S}$; however this is a good approximation to
$f_{\rm esc}$ because the stars are centrally concentrated in its halo
galaxies.

\section{Results}

\subsection{Star Formation and Feedback}
\label{sec:sf}

Here we present our results of our simulations that consider the
self-consistent treatment of star formation and feedback in
high-redshift dwarf galaxies with masses between 10$^{6.5}$ and
10$^{9.5}$ solar masses.  We first discuss the amount and trends with
halo mass of the star formation in these galaxies.  Then we illustrate
the effects of radiative feedback on the gas morphology and
self-regulation of star formation. Last we describe the star formation
histories of such galaxies.

%
%
\begin{figure}[t]
  \begin{center}
    \epsscale{1.15}
    \plotone{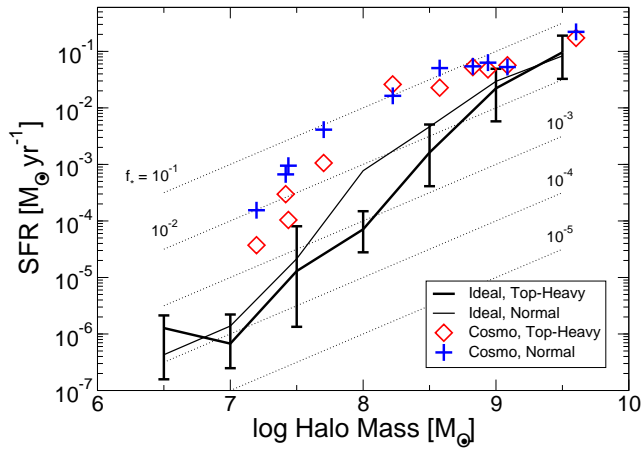}
    \caption{\label{fig:sfr_mass} The mean of the time-averaged star
      formation rates (SFR) for idealized halos.  The thin and thick
      lines correspond to runs with a normal ($N_\gamma = 2,600$) and
      top-heavy ($N_\gamma = 26,000$) IMF, respectively.  The error
      bars show the minimum and maximum values at each halo mass.  The
      average SFR for cosmological halos with a top-heavy (normal) IMF
      are represented with diamonds (plus signs).  The dotted lines show
      the SFR for a constant star formation efficiency, $f_\star =
      M_\star / M_{\rm{gas}}$, for $f_b$ = 0.1 and lasting for 100
      Myr.}
  \end{center}
\end{figure}
%
%
\begin{figure}[t]
  \begin{center}
    \epsscale{1.15}
    \plotone{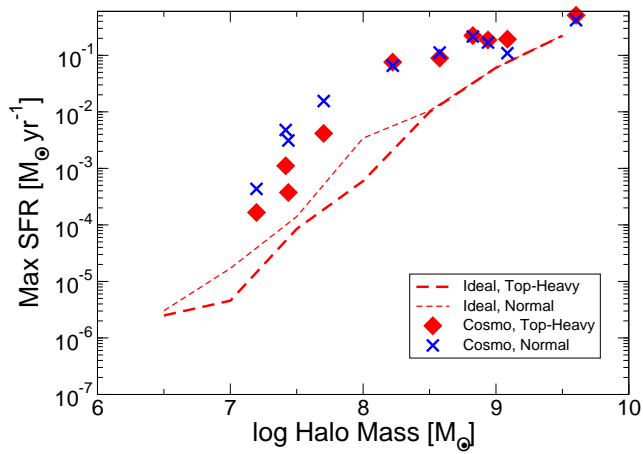}
    \caption{\label{fig:maxsfr_mass} The maximum SFR for idealized
      halos with a normal (thin dashed) and top-heavy IMF (thick
      dashed) and cosmological halos with a normal (crosses) and
      top-heavy IMF (diamonds).}
  \end{center}
\end{figure}
%
%
\begin{figure}[t]
  \begin{center}
    \epsscale{1.15}
    \plotone{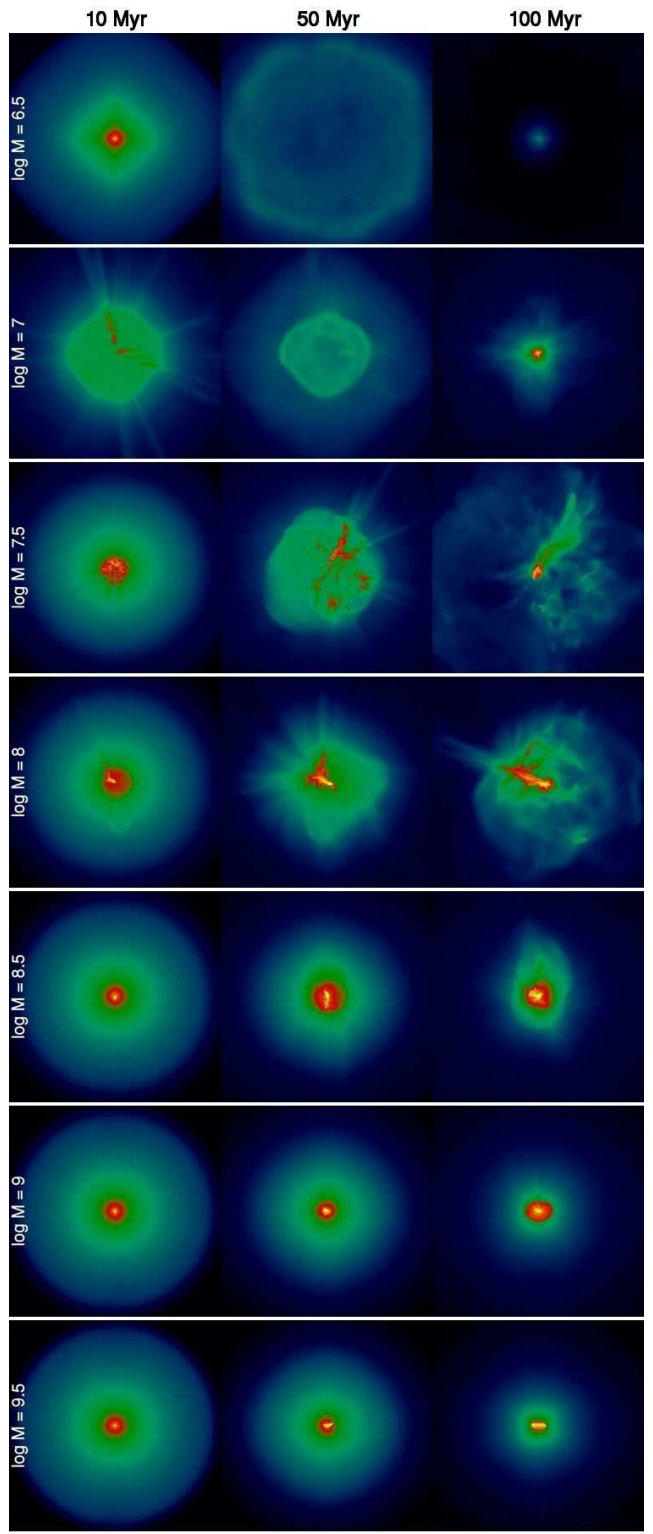}
    \caption{\label{fig:density_evo_iso} Gas density projections,
      weighted by density squared for idealized halos.  The field of
      view is 2\rvir.  The colormap spans 5 orders of magnitude from
      \tento{-3} to \tento{2} cm$^{-3}$.}
  \end{center}
\end{figure}
%
%
\begin{figure*}[t]
  \begin{center}
    \epsscale{1.15}
    \plottwo{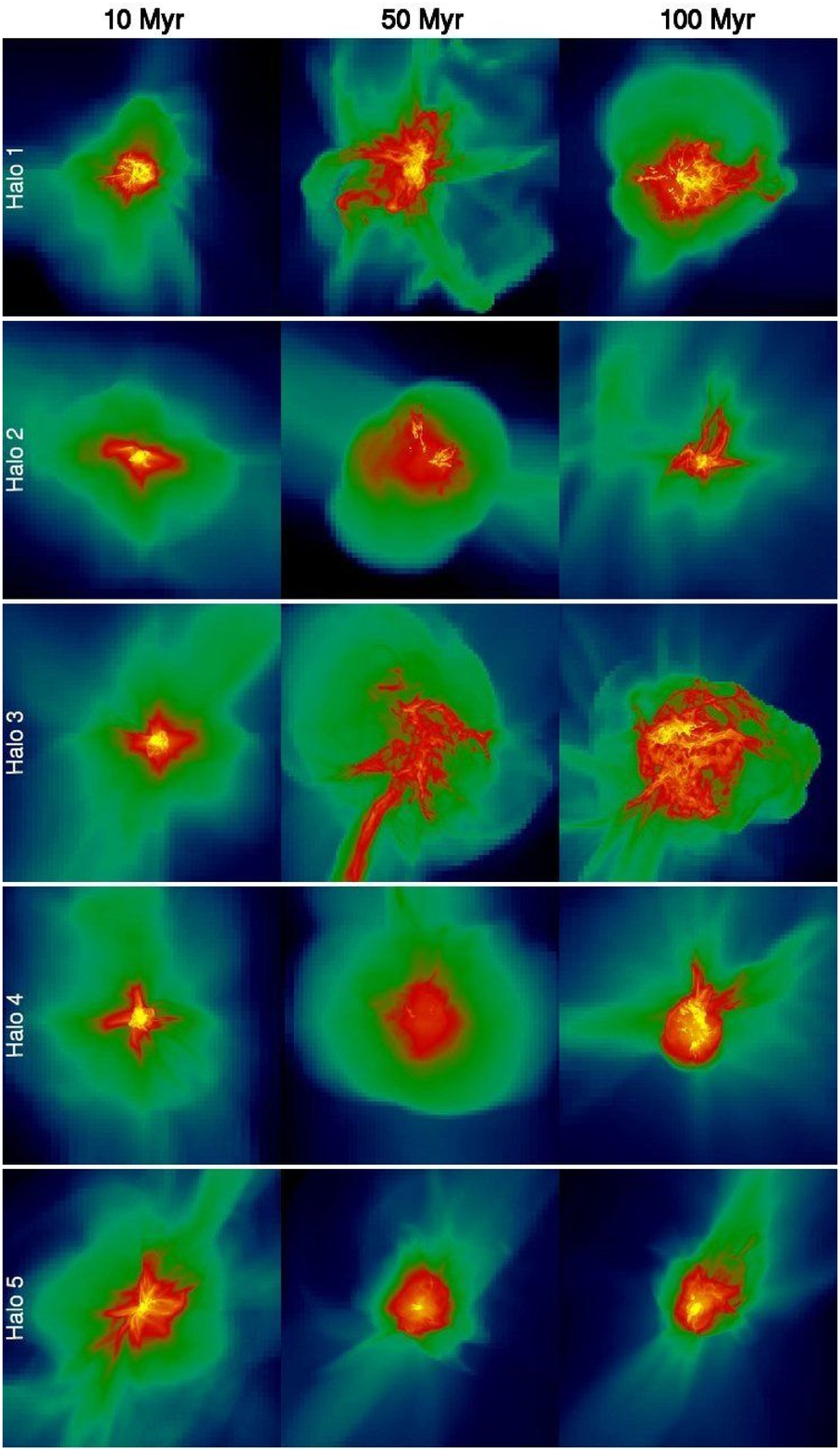}{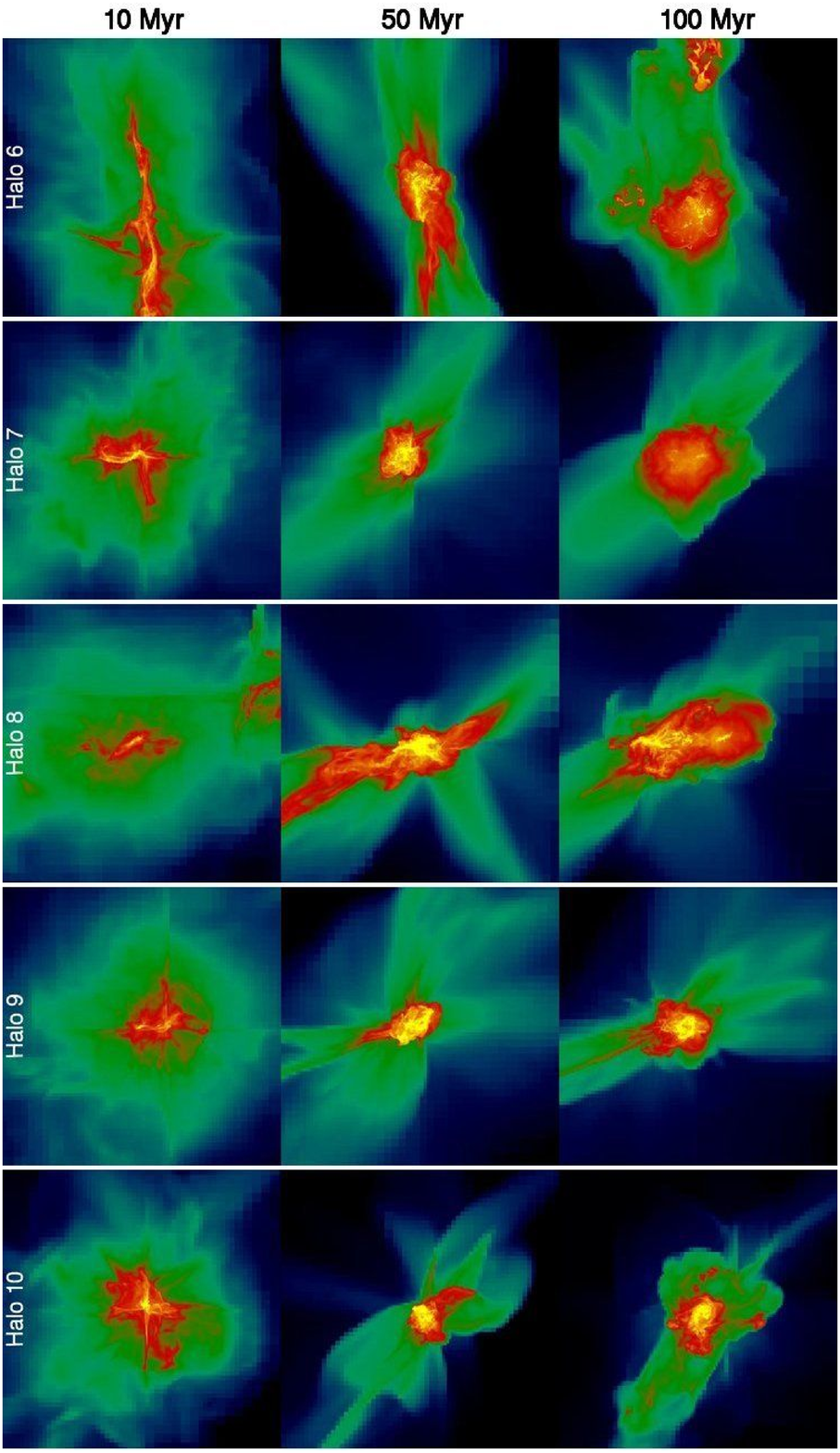}
    \caption{\label{fig:density_evo} Gas density projections, weighted
      by density squared for cosmological halos.  Halos 1--10 have
      $\log_{10}(\mvir)$ = 7.2, 7.4, 7.4, 7.7, 8.2, 8.6, 8.8, 8.9,
      9.1, and 9.6.  The field of view is 2\rvir.  The color scaling
      is the same as Figure \ref{fig:density_evo_iso}.}
  \end{center}
\end{figure*}

\subsubsection{Star Formation Rates}
\label{sec:sfr}

The escape fraction of UV radiation from halo is primarily dependent
on the strength of the intrinsic ionizing luminosity and how it
propagates through the gas into the IGM.  The halos undergo an initial
collapse and ensuing starburst.  Afterwards star formation is
self-regulated by radiative and supernovae feedback.  We find that
most of the UV radiation escapes during the most intense episodes of
star formation, thus an important quantity to investigate is the
maximum SFR of the halo, as well as the average.  We plot the
time-averaged and maximum SFR of the idealized and cosmological halos
with a top-heavy and normal IMF in Figures \ref{fig:sfr_mass} and
\ref{fig:maxsfr_mass}, respectively.  We also plot SFRs if a constant
star formation efficiency, $f_\star = M_\star / M_{\rm{gas}}$, is
assumed for a baryon mass fraction of 0.1 and a starburst lasting for
100 Myr.

The maximum SFRs are anywhere between 2--10 times the average SFR.
The error bars show variances up to an order of magnitude in the star
formation for a given halo mass in the idealized runs.  For halos with
$\mvir > 10^7 \Ms$, SFRs steadily increase as SFR $\propto
\mvir^\alpha$ with $\alpha \sim 2$.  Idealized halos with \mvir~=
\tento{6.5}~and \tento{7} \Ms~have similar SFRs because they
experience only one episode of star formation that blows away most of
the gas in the halo, quenching any further star formation.  In a more
realistic cosmological setup, these outflows only delay later star
formation because the gas reservoir can be replenished through gas
accretion from filaments and mergers \citep{Wise08a}.  This accretion
causes the SFRs to be higher up to a factor of 100 in cosmological
halos up to $\mvir = 10^{8.5} \Ms$, which can be seen in Figures
\ref{fig:sfr_mass} and \ref{fig:maxsfr_mass}.  Above this halo mass,
SFRs interestingly level off at $\sim$0.1 \Ms~yr$^{-1}$, corresponding
to an average star formation efficiency $f_\star = 0.05-0.1$.

With a normal IMF, average and maximum SFRs in cosmological halos with
$\mvir \le 10^8 \Ms$ increase by an order of magnitude.  In these
halos, outflows generated by radiative feedback is the main force in
regulating SFRs.  Therefore one expects less self-regulation of star
formation, and thus higher SFRs, with a normal IMF when compared to a
top-heavy one.  In more massive halos, only outflows generated by SN
explosions can escape from the potential well.  Hence decreasing the
specific stellar luminosity does not significantly affect the gas mass
inside the halo, leading to similar SFRs in the top-heavy and normal
IMF runs.

%
%
\begin{figure*}[t]
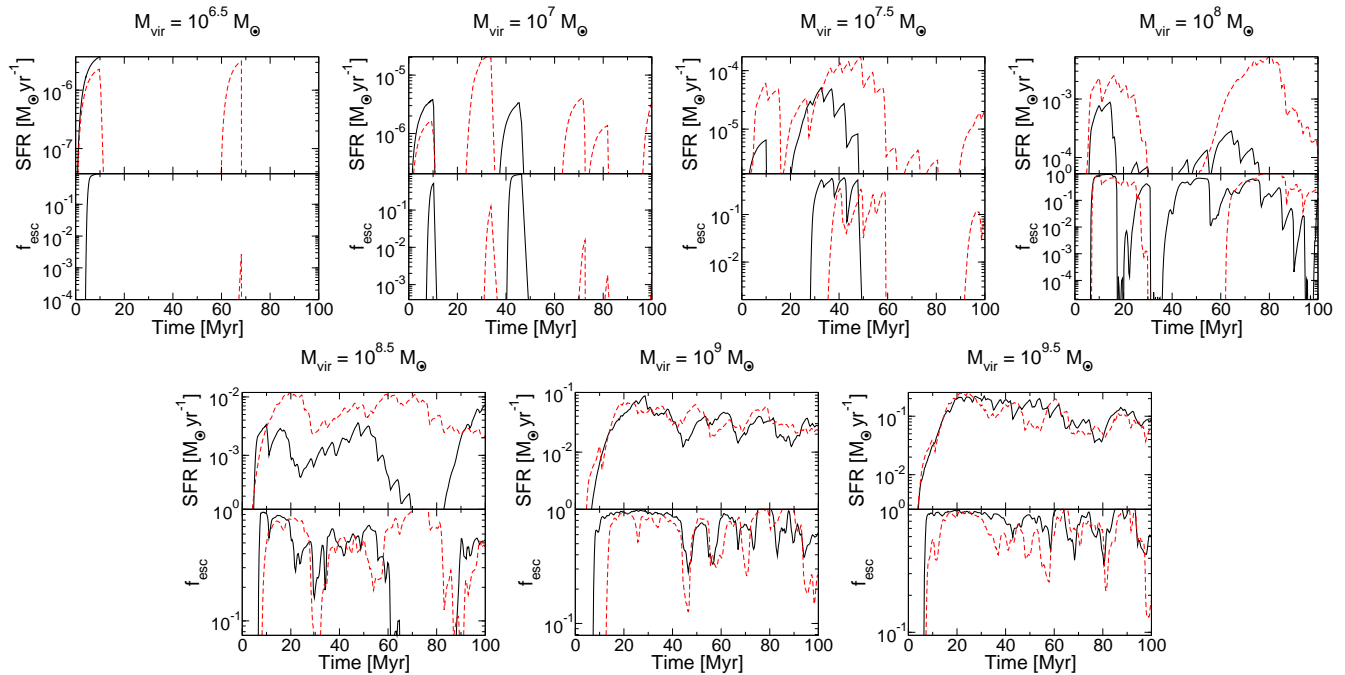

  \begin{center}
    \includegraphics[width=0.24\textwidth]{f5a_color}
    \includegraphics[width=0.24\textwidth]{f5b_color}
    \includegraphics[width=0.24\textwidth]{f5c_color}
    \includegraphics[width=0.24\textwidth]{f5d_color}
    \includegraphics[width=0.24\textwidth]{f5e_color}
    \includegraphics[width=0.24\textwidth]{f5f_color}
    \includegraphics[width=0.24\textwidth]{f5g_color}
    \caption{\label{fig:history1} Star formation rates (top panels)
    and UV escape fractions (bottom panels) for the idealized halo
    setup with $N_\gamma = 26,000$ (solid) and 2,600 (dashed).  The
    data are from models with ($f_b$, $\lambda$, $f_{\rm{turb}}$) =
    (0.1, 0.04, 0.25).}
  \end{center}
\end{figure*}

%
%
\begin{figure*}[t]
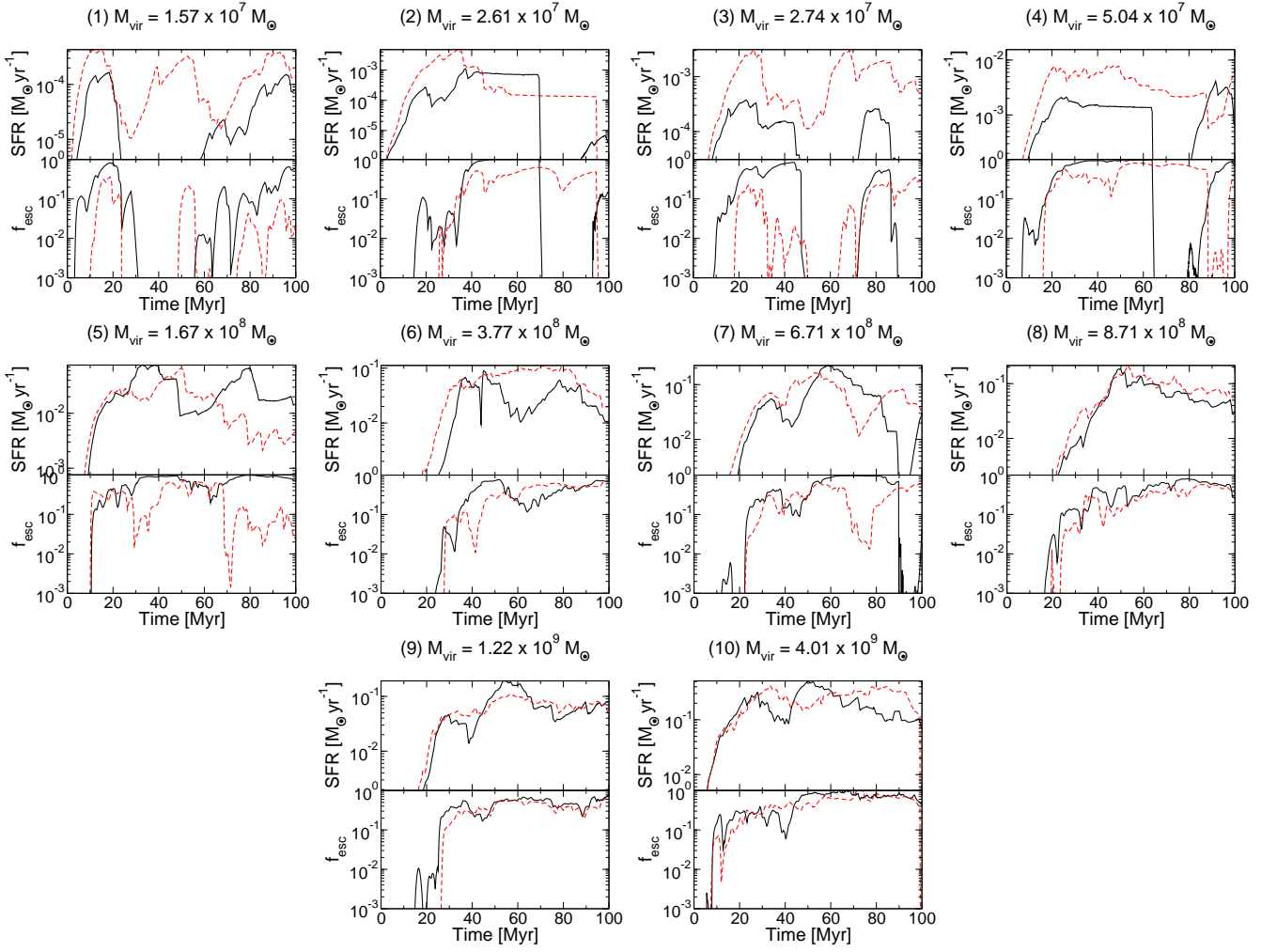

  \begin{center}
    \includegraphics[width=0.24\textwidth]{f6a_color}
    \includegraphics[width=0.24\textwidth]{f6b_color}
    \includegraphics[width=0.24\textwidth]{f6c_color}
    \includegraphics[width=0.24\textwidth]{f6d_color}
    \includegraphics[width=0.24\textwidth]{f6e_color}
    \includegraphics[width=0.24\textwidth]{f6f_color}
    \includegraphics[width=0.24\textwidth]{f6g_color}
    \includegraphics[width=0.24\textwidth]{f6h_color}
    \includegraphics[width=0.24\textwidth]{f6i_color}
    \includegraphics[width=0.24\textwidth]{f6j_color}
    \caption{\label{fig:history2} Same as Figure \ref{fig:history1}
      but for cosmological halos.  The halo numbers are labeled in
      parentheses.}
  \end{center}
\end{figure*}

\subsubsection{Stellar Feedback}
\label{sec:feedback}

In the dwarf galaxies that we consider, stellar feedback from
\ion{H}{2} regions and SN explosions creates galactic outflows that
expel significant amounts of gas from the halo.  Our simulations
capture the formation of a D-type ionization front that accelerates
matter up to 35 \kms~\citep[e.g.][]{Franco90, Whalen04, Kitayama04,
  Abel07}, which is greater than the escape velocities of halos with
$\mvir \lsim 10^{8.5} \Ms$ at redshift 8.  Above this mass, SNe feedback,
which generates outflows up to $\sim$300\kms, becomes the dominant
force in expelling gas.  Feedback strongly affects the gas structures
within the dwarf galaxies, which we illustrate in density-squared
projections of gas density of idealized and cosmological halos in
Figures \ref{fig:density_evo_iso} and \ref{fig:density_evo} at 10, 50,
and 100 Myr after the initial starburst.

The transition from complete gas ``blow-out'' to ``blow-away'' is
apparent in the idealized halos (Fig. \ref{fig:density_evo_iso}) at $M
\gsim 10^{8.5} \Ms$.  In halos with $\mvir = 10^{6.5-7} \Ms$, the
D-type front is unbound from the halo and expels most of the gas with
it.  Only a small fraction returns to the halo center after 100 Myr.
In halos with $\mvir = 10^{7.5-8} \Ms$, the blow-out becomes less
effective at completely evacuating the halo because of the deeper
potential well.  As stated before when the D-type front velocity is
less than the escape velocity of the halo, radiative feedback cannot
create unbound galactic outflows.  Nevertheless, it still plays a key
role in shaping the small-scale ($l \lsim 100$~pc) variations in the
ISM density and temperatures.

The cosmological halos provide a more realistic environment and halo
setup---inflows from filaments, minor mergers, turbulent velocity
fields, an evolving DM potential---in studying the effects of star
formation and feedback.  Despite the more realistic gas inflows in
such halos, we find a similar transition from blow-out to blow-away at
$\mvir = 10^{8.5} \Ms$.  The halos with $\mvir < 10^8 \Ms$ are not
evacuated like their idealized counterparts because of additional gas
accretion from adjacent filamentary structures and minor halo mergers
\footnote{In even less massive halos ($\mvir \sim 10^6 \Ms$), Pop III
  radiative feedback can completely evacuate its host halo in a
  cosmological environment \citep{Abel07, Yoshida07, Johnson07,
    Wise08a}.}.  In response, star formation is self-regulating, in
which the stellar radiation reduces the amount of cool and dense gas
available for star formation.  In the least massive halos, star
formation ceases for tens of millions of years as the relic \ion{H}{2}
region cools and re-condensed.  We further discuss star formation
rates in \S\ref{sec:sfhistory1}--\ref{sec:sfhistory2}.

Above \tento{8} \Ms, radiative feedback does not destroy all of the
density enhancements, in particular one centred in the halo center,
and stars continue forming in the presence of nearby star clusters.
The infalling subhalo ($\mvir \sim 10^7 \Ms$) in Halo 6 at $t$ = 100 Myr
provides an interesting contrast in the halo mass dependence of
radiative feedback, where the main halo is centrally concentrated while
the subhalo is being blown apart by radiative feedback.  In these more
massive halos, star formation is still self-regulated but not
suppressed as there is a persistent cold gas reservoir available for
star formation, as seen in the evolution of gas density in the right
column of Figure \ref{fig:density_evo}.

\subsubsection{Star Formation History of Idealized Halos}
\label{sec:sfhistory1}

We plot the star formation histories of the idealized control halos
($f_b$ = 0.1, $\lambda$ = 0.04, $f_{\rm{turb}}$ = 0.25) in the top
panels of Figure \ref{fig:history1}.  Star formation in halo masses
below \tento{8}\Ms~is episodic.  This occurs when radiation driven
stellar outflows expel most of the gas from the halo, quenching star
formation.  Once the gas is no longer irradiated, it can cool and
fallback into the halo center, forming stars once again.  This occurs
up to 5 times in these halos in the 100 Myr, more often in more
massive halos.

When halo masses increase to between \tento{7.5} and \tento{8.5} \Ms,
star formation continues in the presence of galactic outflows; however
it soon halts once the cool gas reservoir in the halo is depleted.
In Figure \ref{fig:history1}, one sees in halos with these masses
experience at least one period of quiescence, where a substantial
fraction of gas cannot form star-forming molecular clouds because it
is contained in outflows.  Similar to the lower-mass halos, star
formation recommences after gas falls back into the halo.

Above \tento{9}\Ms, the bursting behavior of star formation is
entirely eliminated.  The potential well of the halo has become deep
enough ($V_c > 35$ \kms) to contain any outflows created by
over-pressurized \ion{H}{2} regions.  The transition circular velocity
from baryon ``blow-out'' to ``blow-away'' is similar to the velocities
generated in D-type ionization fronts.  Furthermore,
unlike the lower mass halos, 
the roughly constant SFRs are approximately the same between the
high- and low-luminosity models at $5 \times 10^{-2}$ and
$10^{-1}$~\Ms~yr$^{-1}$ for $M_{vir} = 10^9$ and $10^{9.5} \Ms$,
respectively.

Thin disk formation is absent in halos that experience baryon blowout
because the outward motions created in \ion{H}{2} regions disrupt any
global organized rotation.  In halos with $\mvir > 10^9 \Ms$, the
majority of gas is retained within the halo, which then settles into a
thin disk.  We caution that is probably not strong evidence for a
transition from a globally turbulent ISM to disk formation because of
our idealized setup of solid-body rotation.  We can better investigate
any morphological trends in cosmological halos that are described in
the following section.

\subsubsection{Star Formation History of Cosmological Halos}
\label{sec:sfhistory2}

We show the star formation histories for the 10 selected cosmological
halos in the top panels of Figure \ref{fig:history2}.  The SFR in
cosmological halos is less episodic than the idealized halos because
the cosmological gas accretion replenishes the galaxy, balancing the
effect of any blow-away.  Nevertheless star formation is still
self-regulated, which can be seen in the SFR being suppressed up to an
order of magnitude after a local maximum in SFR.  Afterwards the SFR
recovers again because of the reduced radiative feedback.  This
modulation occurs in all cosmological halos studied here.

In lower mass halos with $\mvir < 10^8 \Ms$ with a top-heavy IMF, star
formation is still completely eliminated for 10--20 Myr.  This does
not occur in the same halos with a normal IMF because the galactic
outflows are weaker, making it easier for the cosmological accretion
to penetrate through the outflows.  Hence star formation is continuous
in these cases.  Similar to the $\mvir > 10^9 \Ms$ idealized halos, SFRs
are not significantly affected by a lower specific stellar luminosity
in cosmological halos with $\mvir > 10^8 \Ms$.  There are some deviations
by a factor of a few in the evolution of the SFR in these halos, but
the time-averaged rates are largely unaffected.  One clear example of
this behavior occurs in Halo 6 ($\mvir = 3.8 \times 10^8 \Ms$) at t =
70 Myr.  In the top-heavy IMF case, radiative feedback from the initial
burst subdues star formation from t = 50--80 Myr by a factor of 5.  In
contrast, the feedback from a normal IMF does not suppress star
formation, which continues at a rate of $\sim$$10^{-1}
\Ms~\rm{yr}^{-1}$.  Furthermore, galactic disk formation, which is
seen in isolated halos with $\mvir \ge 10^9 \Ms$, is absent in all of
the cosmological cases.

\subsubsection{Comparison to Previous Work}

The SFRs in the $\mvir = 10^{9.5} \Ms$ halos are
$\sim$\tento{-1}~\Ms~yr$^{-1}$ and are similar to the least massive
(\mvir~$\sim$ \tento{10}\Ms) halos in GKC08.  However for a direct
comparison of SFRs and escape fractions, we must either use the
simulations with $N_\gamma$ = 2,600 or consider the SFRs from
top-heavy IMF simulations as effectively 10 times greater.
\citet{Fujita03} characterizes star formation with a star formation
efficiency $f_\star = M_\star / M_d$, where $M_d$ is the total gas
disk mass.  If we equate the total gas mass in our halos to the disk
mass of \citeauthor{Fujita03}, our $\mvir = 10^9 \Ms$ idealized halos
are comparable to their high-redshift $M_d = 10^8 \Ms$ models with
$f_\star = 0.06$.

\subsection{UV Escape Fraction}
\label{sec:fesc}

A fraction of the stellar radiation emitted from the stars described
in the previous section escape into the IGM.  This quantity is an
essential parameter in both semi-analytical reionization models
\citep[e.g.][]{Cen03a, Cen03b, Wyithe03a, Haiman03} and N-body
reionization simulations \citep[e.g.][]{Ciardi03, Sokasian03, Iliev06,
  Trac07}.  First we discuss the trends of $f_{\rm esc}$ with halo mass of
the idealized halos.  We then describe the behavior of $f_{\rm esc}$ with
respect to important physical parameters: turbulent energy, halo spin
parameter, and baryon mass fraction.  Next we discuss the results of
$f_{\rm esc}$ in halos that have been extracted from cosmological
simulations.  We end this section discussing anisotropic \ion{H}{2}
regions and its relevance in the UV escape fraction.

%
%
\begin{figure}[t]
  \begin{center}
    \epsscale{1.15}
    \plotone{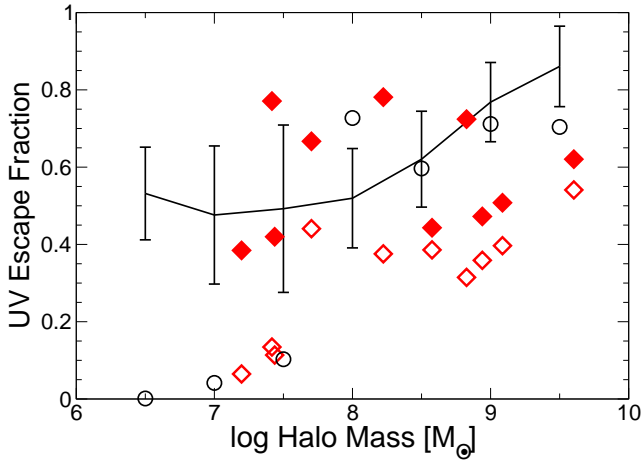}
    \caption{\label{fig:fesc_mass} Escape fraction of ionizing photons
      from idealized halos with a top-heavy IMF (solid line) and
      normal IMF (circles) and cosmological halos with a top-heavy IMF
      (filled diamonds) and normal IMF (open diamonds) as a function
      of halo mass.}
  \end{center}
\end{figure}

\subsubsection{Idealized halos}

We plot the evolution of $f_{\rm esc}$ of the idealized halos in the
bottom panels of Figure \ref{fig:history1}.  Before presenting our
results of $f_{\rm esc}$ in these halos, we would like to warn that
one should not take the values of $f_{\rm esc}$ as absolute because of
the lack of gas accretion and adjacent filamentary structures.
Nevertheless it is appropriate to study the temporal variations and
changes with physical parameters.

In halos with $\mvir \le 10^{7} \Ms$, $f_{\rm esc}$ behaves in a
predictable manner, in which star formation is confined within a small
region and happens in bursts.  Thus most of the escaping radiation
experiences similar gaseous structures when propagating outwards.  In
the bursts, the $f_{\rm esc}$ quickly increases when the \ion{H}{2}
region breaks out from the halo, which can take up 10 Myr after the
starburst starts.  The escape fraction peaks when the SFR is at its
maximum and varies anywhere from \tento{-3} to unity, depending on the
column density of the surrounding medium.

In more massive halos, the escape fraction is highly dependent on
preceding star formation and its effect on the gas structure.  For
example in halos with $\mvir$ = \tento{7.5} -- \tento{8.5} \Ms, SFRs
that differ by an order of magnitude can produce similar values of
$f_{\rm esc}$.  In the initial starburst, the radiation needs to
escape through a relatively smooth $r^{-2}$ density profile and
$f_{\rm esc} \ge 0.5$.  Then conditions quickly erode from this smooth
initial state to a turbulent multi-phase ISM because of both radiative
cooling and stellar feedback.  Now radiation must escape through a
clumpy and turbulent ISM, whose porosity can increase \fesc~for a
given luminosity or SFR \citep{Clarke02}.

Generally only radiation from the strongest starbursts can escape,
whereas during suppressed periods of star formation very little
($f_{\rm esc} < 0.01$) radiation escapes.  As SFRs become more
consistent in halos with $\mvir > 10^9 \Ms$, the escape fraction
varies between 0.1 and 1.0 with less variation than the lower mass
halos.  In addition, we find that radiation from stars that form on
the outskirts of the thin disk have a higher escape fraction than
stars that are deeply embedded within the galaxy, which is similar to
the results of GKC08.

%
%
\begin{figure}[t]
  \begin{center}
    \epsscale{1.15}
    \plotone{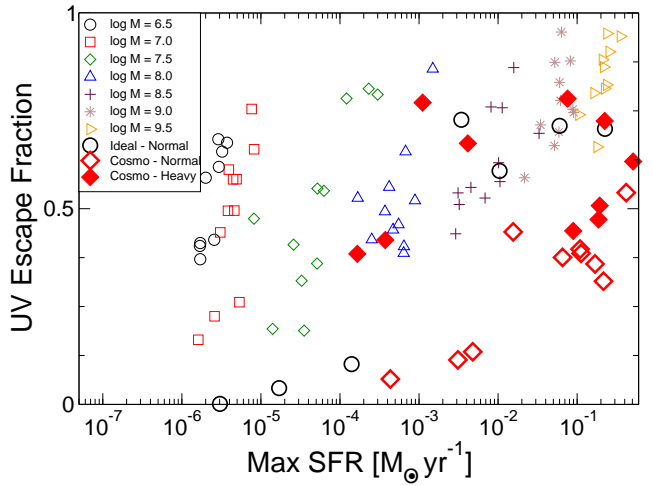}
    \caption{\label{fig:fesc_sfr} Escape fraction of ionizing photons
      as a function of maximum star formation rate from all simulated
      halos, showing a strong dependence for a given halo mass.
      Idealized halos with equal halo masses are plotted with the same
      symbols.  Idealized halos with a normal IMF are plotted as large
      circles ranging from $10^{6.5}$ to $10^{9.5} \Ms$ from left to
      right.  Cosmological halos with a top-heavy and normal IMF are
      plotted as filled and open diamonds, respectively.}
  \end{center}
\end{figure}

In Figure \ref{fig:fesc_mass}, we plot the average UV escape fraction
as a function of halo mass, where the error bars indicate the standard
deviation of $f_{\rm esc}$ over the 11 models for each halo mass.
Halos with $\mvir \le 10^8 \Ms$ have $f_{\rm esc} \sim 0.5 \pm 0.2$.
For a normal IMF in these low-mass halos, \fesc~drops to values below
0.1.  Escape fractions steadily increase with halo mass to $\sim$0.8
above this mass for both top-heavy and normal IMFs.  

As seen in Figure \ref{fig:history1}, $f_{\rm esc}$ is highest when
the SFR is at its maximum, max(SFR).  To depict this strong dependence
on the strength of the starburst, we show $f_{\rm esc}$ for every
simulated halo as a function of max(SFR) in Figure \ref{fig:fesc_sfr},
grouped by halo mass.  For instance in the $\mvir = 10^{7.5} \Ms$
halos, $f_{\rm esc}$ increases from 0.2 to 0.8 when max(SFR) is higher
by an order of magnitude.  We have also plotted the idealized halos
with a normal IMF as large open circles.

\subsubsection{Dependence on halo parameters}
\label{sec:trends}

\begin{figure*}[t]
  \begin{center}
    \epsscale{1.0}
    \plotone{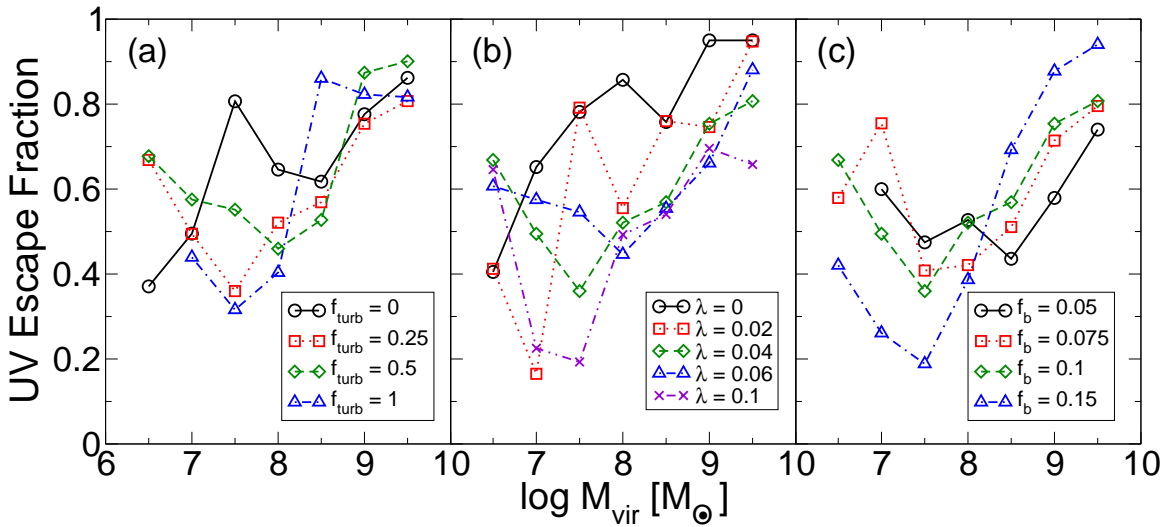}
    \caption{\label{fig:trends} UV escape fraction versus halo mass
      plotted for varying values of (a) turbulent energy, (b) spin
      parameter, and (c) baryon mass fraction.}
  \end{center}
\end{figure*}

Although the absolute values of $f_{\rm esc}$ in idealized halos are
inaccurate for reasons discussed in the previous section, we can
utilize the trends with physical halo parameters---baryon mass
fraction, spin parameter, and turbulent energy---to estimate $f_{\rm esc}$
in cosmological halos other than the ones studied in this paper.  For
example, \citet{Wise08a} found that the first stars lowered the baryon
mass fraction to $\le 0.1$, but the halos from our adiabatic
cosmological simulations have $f_b \sim 0.13$.  We can then use the
trends in $f_{\rm esc}$ with respect to $f_b$ to estimate the effects of
neglected processes on the escape fraction.  We show these trends for
all idealized halos in Figure \ref{fig:trends}.

%
%

%
%
\begin{figure*}[t]
  \begin{center}
    \epsscale{1.11}
    \plottwo{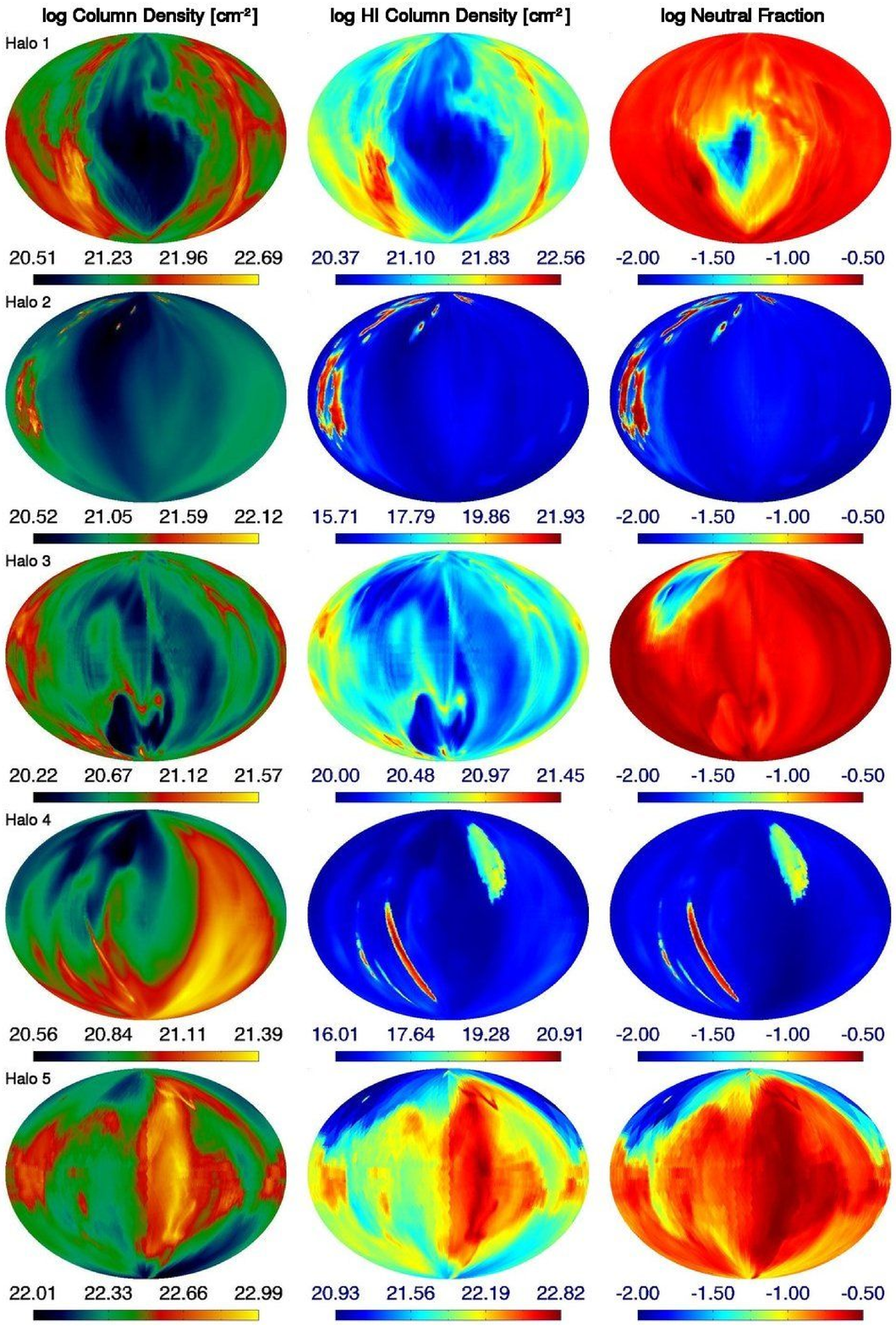}{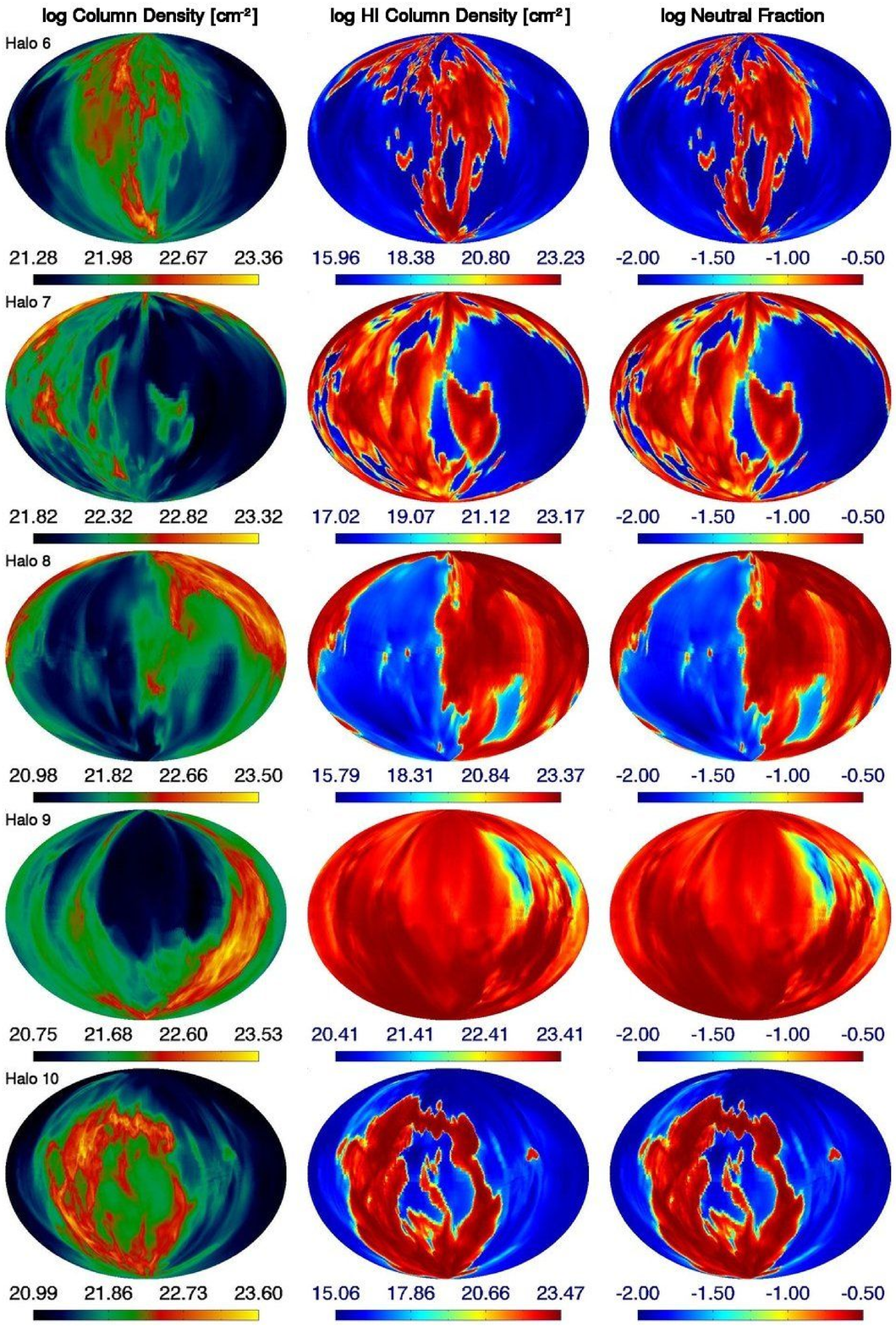}
    \caption{\label{fig:maps} Full-sky maps of total column density
      (first and fourth columns), \ion{H}{1} column density (second
      and fifth columns), and mass-weighted neutral hydrogen fraction
      (third and sixth columns) 50 Myr after the initial starburst in
      the cosmological halos.  Halos 1--10 have $\log_{10}(\mvir)$ =
      7.2, 7.4, 7.4, 7.7, 8.2, 8.6, 8.8, 8.9, 9.1, and 9.6.  The
      observer is at the center of mass of the halo, and the
      quantities shown out to the virial radius.  The colormaps for
      column density are unique for each halo; however the electron
      fraction colormaps have a fixed range of 0.01--0.3.  Halos 2 and
      4 are almost completely ionized, and halos 6, 7, and 10
      represent clear examples of anisotropic \ion{H}{2} regions,
      which escape from the halo where the column density is lowest.}
  \end{center}
\end{figure*}

\medskip

\textit{Turbulent energy}--- Initial turbulent energy should be
unimportant in the more massive halos as radiative and SN feedback
continuously alters the ISM and further stirs turbulence
\citep[e.g.][]{Scalo04}, perhaps overpowering any pre-existing
turbulence.  There are no consistent trends with turbulent energy
between \tento{6.5} -- \tento{9.5} \Ms.  However in the $\mvir = 10^{7.5}
\Ms$ halos, the $f_{turb} = 0$ case has $f_{\rm esc} = 0.8$ that is up to
twice the escape fraction of the other halos.  This occurs because
turbulent flows are not present to disrupt the initial halo collapse,
resulting in a stronger initial starburst and escalating the average
escape fraction.  The $f_{turb} = 0$ case with $\mvir = 10^8 \Ms$ is only
marginally higher than the other halos by $\sim$25\%.  This anomaly is
not present in other halo masses.


\medskip

\textit{Spin parameter}--- Similar to the lack of apparent trends with
turbulent energy, spin parameter has little effect on the UV escape
fraction.  One exception is the non-spinning case above \tento{7} \Ms,
having the largest escape fraction.  This arises from a strong central
collapse without being rotationally supporting, resulting in a
stronger initial starburst.  However this scenario is improbable in
reality because of abundant cosmological tidal torques \citep{Hoyle49,
  Peebles69}, culminating in a log-normal spin parameter distribution
that has a mean value of 0.042 \citep[e.g.][]{Barnes87, Bullock01a}.

\medskip

\textit{Baryon mass fraction}--- Out of the three physical parameters
studied, the baryon mass fraction plays the largest role in regulating
star formation and the escaping radiation.  In halos with $\mvir < 10^8
\Ms$, the $f_b = 0.15$ case has $f_{\rm esc}$ lower than the other gas
fractions by a factor of 2.  The increased neutral column density that
the radiation must transverse before breaking out of the halo causes
the decrease.  Interestingly above \tento{8} \Ms, this trend reverses,
and higher gas fractions result in larger escape fractions.  Here the
gas-rich halos can sustain a higher SFR, and the corresponding higher
ionizing luminosity can escape the halo more freely than in the
gas-poor ($f_b \le 0.1$) halos, even though the radiation has a larger
column to ionize.

\subsubsection{Cosmological halos}
\label{sec:fesc_cosmo}

The star-forming cosmological halos represent a more realistic setup
of high-redshift, starburst dwarf galaxies, necessary to calculate a
robust value of the UV escape fraction.  In the lower panels of Figure
\ref{fig:history2}, we show the escape fraction of the cosmological
halos as a function of time.  During the first starburst that
originates from the initial halo collapse, instantaneous $f_{\rm esc}$
values widely vary from less than 10$^{-3}$ up to nearly unity.  Most
importantly the escape fraction depends on the current star formation
and any of its time variations.  As seen in Figure \ref{fig:history2}
the variations in SFR corresponds to variations in $f_{\rm esc}$ on
similar timescales, on order of a dynamical timescale of a molecular
cloud, $\sim$1--3 Myr.  In halos 2 and 4, a constant SFR exists for 30
Myr in the top-heavy IMF case.  The cumulative luminosity from this
stellar cluster gradually ionizes the IGM and photo-evaporates most
clumpy material.  Eventually the escape fraction approaches unity
after $\sim$10 Myr of irradiation.

As with the idealized halos, the escape fraction is dependent on the
previous star formation history, which can photo-evaporate dense,
clumpy material in the halo.  Imagine two equal-mass stellar clusters,
forming in the same neighborhood.  The one that forms first will
pre-ionize the ISM in some fraction of solid angle, allowing the
radiation from the second cluster to escape into the IGM more easily
and further raising the escape fraction.

We compare $f_{\rm esc}$ from cosmological halos with their idealized
counterparts in Figures \ref{fig:fesc_mass} and \ref{fig:fesc_sfr}
with respect to virial mass and maximum SFR, respectively.  The halos
that host a top-heavy IMF have escape fractions within the scatter of
the idealized halo, even though adjacent filamentary structures can
absorb most of the radiation in their line of sight.  Their $f_{\rm
  esc}$ values range from 0.4 to 0.8.  We also find that \fesc~at half
and twice the virial radius are within 10\% of the value at \rvir,
similar to GKC08, because the nearby clumps and adjacent filaments do
not contribute a significant solid angle for absorption.

A normal IMF lowers the escape fraction by $\Delta(f_{\rm esc})$ =
0.05--0.4.  It is interesting that $f_{\rm esc}$ is approximately 0.4
with a normal IMF over 2 orders of magnitude in halo mass and maximum
SFR, starting with $\mvir = 10^{7.5} \Ms$.  Halos below this mass
threshold (i.e. efficient atomic cooling) are affected considerably by
a normal IMF, where \fesc~drops by a factor of a few to values of
0.05--0.1.

\subsubsection{Anisotropic \ion{H}{2} Regions}

In our radiation hydrodynamics simulations, radiative feedback greatly
affects the gas dynamics inside the halos.  Any UV radiation will
first escape from the halo in the direction with the least \ion{H}{1}
column density.  This creates anisotropic \ion{H}{2} regions with the
radiation preferentially escaping through these channels.  Any
adjacent filamentary structures, which provide the halo with a cold,
dense flow \citep{Nagai03b, Keres05, Dekel06, Wise07a}, and cold and
clumpy ISM are the major components in absorbing outgoing radiation.

We create full-sky maps of total and \ion{H}{1} column density and
mass-averaged neutral fractions of the cosmological halos with a
top-heavy IMF 50 Myr after the initial starburst in Figure
\ref{fig:maps}.  These maps are created by casting adaptive rays
\citep{Abel02b} from the center of mass of the halo to the virial
radius.  Each ray tracks the total and neutral column density along
its path.  We then reconstruct a full-sky map of these quantities
using the spatial information contained in the HEALPix formalism.  The
ionized regions (blue) match well with the areas with the lowest
column density (black and dark blue).  The transitions from neutral to
ionized in these maps are abrupt, depicting how any neutral blobs can
completely absorb the radiation in its solid angle.  However, in the
solid angles where the ionized fraction is close to unity, almost all
of the radiation can escape, i.e. $f_{\rm esc} = 1$ in those
directions.  The simulations of RS06, RS07, and GKC08 also see the
majority of solid angles having $f_{\rm esc}$ equal to either unity or
zero.

At $t = 50$ Myr, halos 1 and 3 are in a quiescent state, and the
previously ionized ISM and IGM is recombining.  The relic \ion{H}{2}
region are still visible in the full-sky maps.  Here the transition
from ionized and neutral regions are gradual because of the longer
recombination times in diffuse gas.  Halos 2 and 4 have been almost
completely ionized at the end of a starburst.  Notice the cometary
tails created by photo-ionization of a clumpy medium
\citep[e.g.][]{Susa06, Whalen08} at the northern section in halo 2 and
the middle-bottom region in halo 4.  In the former case, the neutral
shadow, pointing southwest, of the small overdensities are clear in
the neutral fraction map.  In halos 5--10, stellar radiation escapes
through a large fraction ($>25\%$) of solid angle, which correspond to
the path of least neutral column density.

\subsubsection{Comparison to Previous Work}

Clearly these escape fractions contradict the results from GKC08,
where their lowest mass halos with $\mvir \sim 10^{10} \Ms$ have
$f_{\rm esc} = 10^{-5}-10^{-2}$ at $z = 3-5$.  From the examples given
there, the stars are born within a galactic disk, unlike our
simulations, which could be causing their smaller escape fractions.
In \S\ref{sec:disc}, we further discuss possible causes of this
discrepancy.  Although our galaxies are smaller by a mass factor of
$>$100 than the ones simulated in RS06 and RS07, it is worth
reiterating that they find $f_{\rm esc} = 0.01-0.1$ at redshift 3.
Both groups also find that $f_{\rm esc}$ can significantly vary from
galaxy to galaxy and in different lines of sight to each galaxy.
Escape fractions in \citet{Fujita03} exceed 0.2 in their high-redshift
dwarf galaxies if the starbursts are intense, i.e. their $f_\star \ge
0.06$ models, similar to our $f_{\rm esc}$ values in the most massive
halos.

\vspace{1em}
\section{Discussion}
\label{sec:disc}

The fraction of ionizing radiation that escapes high-redshift galaxies
is an important value to quantify in order to characterize
cosmological reionization.  In this section, we first discuss the
differences and similarities, along with their respective causes,
between our results and previous work.  Next we discuss any
implications and agreements with semi-analytic reionization models.
We last elaborate on possible influences from neglected physical
processes.

\subsection{Possible Physical and Computational Dependencies}

Our results suggest that $f_{\rm esc}$ is higher than 0.1 in such
galaxies, but we must first understand why our $f_{\rm esc}$ values
differ from the more massive galaxies presented in the simulations of
\citet{Razoumov06, Razoumov07} and \citet{Gnedin08}.  Below we discuss
five possible causes.

\textit{Galaxy morphology and environment}--- The galaxies simulated
here exhibit an irregular morphology and did not form a
rotationally-supported disk, whereas the simulations of RS06, RS07,
and GKC08 all studied disk galaxies.  As mentioned in
\citet{Fujita03}, a turbulent and clumpy ISM may allow for radiation
to escape more easily into the IGM.  Compared to a disk configuration,
there are more low-density escape routes, i.e. porosity, for the
radiation because of the clumpy nature of the gas \citep{Clarke02}.
Furthermore, the stellar orbits are not confined within a disk and can
reach the outskirts of the galaxy at apocenter, where gas densities
are much lower than the galactic center and ionizing radiation can
escape into the IGM at a much greater rate.  Perhaps this irregular
morphology is only initially present in dwarf galaxies because
radiative feedback shifts the angular momentum distribution to higher
values \citep{Wise08a}.  A fraction of this material will return to
the galaxy and aid in disk formation in higher mass galaxies, such as
the ones studied in RS06, RS07, and GKC08.  Furthermore in rare halos,
the intersecting filaments tend to be dense and thin, where a halo
with equal mass at lower redshift would be contained in a large
filament \citep{Ocvirk08}.  This redshift dependency on halo
environment could affect \fesc~at $r > r_{\rm vir}$, but should not be
apparent at \rvir.

\textit{Galactic mass}--- GKC08 find that $f_{\rm esc}$ increases as
much as 3 orders of magnitude from a halo mass of \tento{10}~to
\tento{11}\Ms.  The cold \ion{H}{1} disk in their lower mass galaxies
tend to be more vertically extended, which brings about this
precipitous drop.  If this trend continues to lower masses, the values
of $f_{\rm esc}$ would be negligible in our galaxies, but we see no
such continuation.  Our results suggest that the radiative feedback in
dwarf galaxies with $V_c \lsim 35 \kms$ have a profound impact on the
ISM by driving outflows and preventing any disk formation, increasing
\fesc.  Perhaps this circular velocity is a critical turning point
below which $f_{\rm esc}$ is large ($>0.1$) because of dynamical
effects from radiative feedback.  In more massive galaxies, SN
feedback provides the main impetus for driving outflows.

\textit{Star formation rates}--- One can argue that our simulations
only capture the initial collapse and an artificially strong starburst
in high-redshift galaxies; however, after $\sim$20 Myr of star
formation, this collapse is reversed by stellar feedback and SFRs
stabilize.  In addition our most massive galaxy ($\mvir = 4 \times 10^9
\Ms$) has an average SFR of 0.2 \Ms~yr$^{-1}$, which is comparable to
the SFRs found in the $\mvir \sim 10^{10} \Ms$ galaxies in GKC08.  The
star formation efficiencies ($M_\star/M_{\rm{gas}}$) of our galaxies range
from 5--10\%, also agreeing with previous galaxy formation
simulations.  Hence we do not think our treatment of star formation
affects the SFRs and resulting UV escape fraction.

\textit{Stellar IMF}--- We experimented with both normal ($N_\gamma =
2,600$) and top-heavy ($N_\gamma = 26,000$) IMFs and found that a
normal IMF causes $f_{\rm esc}$ values to be lower by 10--75\%.  The
differences in how the gas and ensuing star formation reacts to
radiative feedback most likely causes this spread.  The normal IMF is
similar to the ones used in all previously quoted theoretical works,
and thus we do not expect our choice in IMF to cause our high UV
escape fractions.  One minor shortcoming of our simulations is that we
keep the SN feedback strength equal in our normal and top-heavy IMFs.
Although this helps us localize the cause of any changes to a
different specific luminosity, the strength of SN feedback should also
affect the gas distribution, SFR, and \fesc.

\textit{Resolution}--- We can afford a spatial resolution of $\sim$0.1
physical pc in all of our simulations.  However in larger galaxies, it
becomes increasingly difficult to maintain this resolution, especially
in cosmological simulations.  The highest resolution simulation in
RS06 and RS07 use a gravitational softening length of 330 $h^{-1}$ pc,
and their radiative transfer grid is refined so that 10 particles
occupy each cell.  GKC08 use AMR simulations with a maximum resolution
of 50 physical pc.  \citet{Fujita03} have a fixed resolution of 0.28
pc in their $z = 8, M_d = 10^8 \Ms$ simulation, similar to our
resolution.  The $f_{\rm esc}$ values of our simulations and
\citeauthor{Fujita03} are in agreement.  Regular and giant molecular
clouds have radii of 2--20 and 10--60 pc \citep[see][for a
review]{MacLow04}, respectively, and are usually accounted with
subgrid physics in cosmological simulations.  However our simulations
capture the turbulent and clumpy nature of the ISM, while resolving
such molecular clouds.  As discussed before, the turbulent ISM and its
porosity may play a key role in allowing radiation to escape
\citep{Clarke02}.  In addition, high-resolution simulations with
radiative transfer can model the fragmentation of ionization fronts
\citep[e.g.][]{Fujita03, Whalen08a}, which can further assist in
boosting UV escape fractions.

\subsection{Impact on reionization models}
\label{sec:models}

In reionization models, whether it be based on Press-Schetcher
formalism or N-body simulations, the product $\fescs \equiv f_\star
\times \fesc$ ultimately dictates the absolute number of photons that
escape from each halo into the IGM.  When calibrating these models
against \textit{WMAP} observations \citep{Spergel07, Komatsu08} and
\lya~transmission from $z \sim 6$ quasars \citep[e.g.][]{Bolton07,
  Srbinovsky08}, the values of $\ge 0.01$ are usually required.
Recently, \citeauthor{Srbinovsky08} found that \fesc~must lie within
the range 0.1--0.25 with a best-fit of $f_\star = 0.11$ if all halos
with masses $\gsim 10^9 \Ms$ contribute to reionization.  Furthermore,
they conclude that even the smallest atomic line cooling halos ($\mvir
\gsim 10^8 \Ms$) have $\fesc \sim 0.05$.  Our results support this
scenario where low-luminosity galaxies are the main contributors to
reionization.

\begin{figure}[t]
  \begin{center}
    \epsscale{1.15}
    \plotone{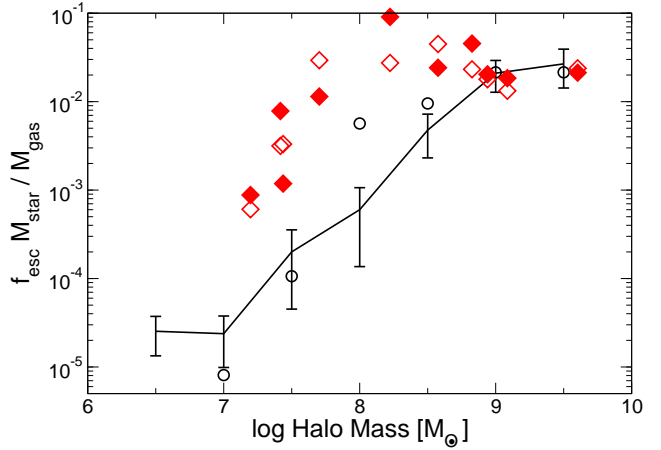}
    \caption{\label{fig:fesc_cstar} The product of stellar mass
      fraction ($M_{\rm star}/M_{\rm gas}$) and escape fraction of
      ionizing photons ($f_{\rm esc}$) from idealized halos and
      cosmological halos with a top-heavy IMF (filled diamonds) and
      normal IMF (open diamonds) as a function of halo mass.
      Idealized halos with a normal IMF are plotted as large circles.
      This product is a key quantity in semi-analytical reionization
      models in determining the amount of escaping radiation per
      collapsed gas fraction.}
  \end{center}
\end{figure}

We plot \fescs~from idealized and cosmological halos in
Figure~\ref{fig:fesc_cstar}.  We calculate $f_\star$ using the initial
gas mass in halo at the initial time; thus we are overestimating
$f_\star$ up to a factor of $\sim$1.5 because the halos experience
mass accretion in the 100 Myr that we have simulated.  Recall that
none of the halos presented here undergo a major merger.  In our least
massive cosmological halos with $\mvir \le 3 \times 10^7 \Ms$,
\fescs~increases rapidly with respect to halo mass from \tento{-3}~to
\tento{-2} because atomic hydrogen line cooling becomes efficient in
these halos.  \fescs~is also an order of magnitude higher than the
idealized cases.  As discussed in \S\ref{sec:sfr}, the additional gas
accretion from filaments results in higher SFRs or equivalently
$f_\star$.  In atomic cooling halos ($\tvir > 8000$ K), we find
\fescs~to always be $\ge 0.01$, sufficiently high enough to meet the
``critical'' value to match reionization constraints, with an average
of 0.033 and 0.026 for top-heavy and normal IMFs, respectively.  When
integrated over a luminosity function with a faint-end slope of --1.7
\citep[e.g.][]{Bouwens07}, the average of \fescs~in atomic cooling
halos is 0.029 and 0.021 for top-heavy and normal IMFs, respectively.
If we include all halos, this average drops to $4.1 \times 10^{-3}$
because the low-mass halos are more abundant and cannot cool and form
stars as efficiently.

\begin{figure*}[t]
  \begin{center}
    \epsscale{1.0}
     \plotone{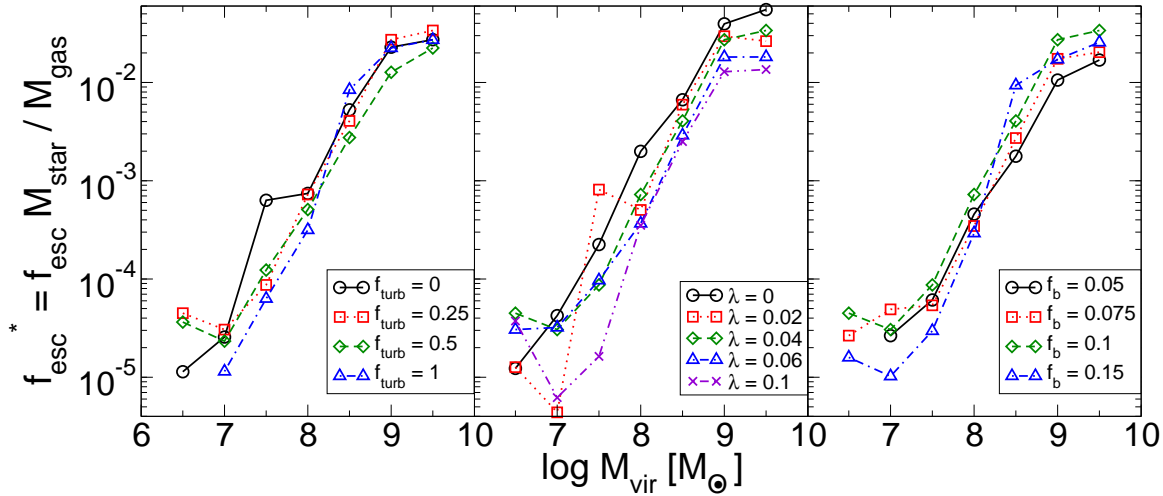}
     \caption{\label{fig:fesc_cstar_trends} Relative changes in the
       product of stellar mass fraction ($M_{\rm star}/M_{\rm gas}$)
       and escape fraction of ionizing photons ($f_{\rm esc}$) from all
       idealized models grouped by initial halo parameters: turbulent
       energy (\textit{left}), spin parameter (\textit{middle}), and
       baryon mass fraction (\textit{right}) when compared to the
       ``control halos'' with ($f_{\rm turb}, \lambda, f_b$) = (0.25,
       0.04, 0.1).}
  \end{center}
\end{figure*}

In Figure~\ref{fig:fesc_cstar_trends}, we show the behavior of
\fescs~with halo mass in idealized halos grouped by halo parameters,
similar to Figure~\ref{fig:trends}.  There is no apparent trends with
respect to turbulent energy; however, halos with higher spin
parameters result in smaller values of \fescs~in high mass halos that
form a rotationally supported disk.  The halos with $f_b = 0.1$ have
the highest values of \fescs, and gas-poor halos are consistently
lower in all halo masses.

%
%
%
%

\subsection{Effects from the Ultraviolet Background}

Our simulations accurately track the evolution of the \ion{H}{2}
regions and how they breakout into the IGM.  Here we discuss an
important process that was neglected in our simulations, 
external feedback from the UV background (UVB).

Our models sample halos that are primarily dependent on \hh~cooling
($\tvir \lsim 8000$ K) and atomic line cooling in more massive halos.
Gas condensation in the lower mass \hh~cooling halos can be delayed by
a \hh~dissociating UV (Lyman-Werner) background between 11.2 and 13.6
eV \citep{Machacek01, Wise07b, OShea08}.  The Lyman-Werner background
thus increases cooling times in the centers of such halos.  As a
result, the minimum mass of a star-forming halo increases with the
Lyman-Werner background intensity.  This will not affect the global
halo properties but may suppress the SFR in these halos.  The
Lyman-Werner background becomes less of an issue in atomic line
cooling halos as \lya~cooling provides ample amounts of free electrons
for \hh~cooling, and they become self-shielding to this radiation
\citep{Wise08a}.

We now consider any photo-heating from a hydrogen ionizing UVB, which
can partially suppress gas accretion and thus star formation in dwarf
galaxies \citep[e.g.][]{Efstathiou92, Shapiro94, Thoul96}.  Here the
Jeans ``filtering mass'' \citep{Gnedin98} accurately describes the
minimum halo mass that can cool and collapse given an IGM thermal
history \citep{Gnedin00, Wise08b}.  The Jeans filtering mass is
calculated by analyzing the linear evolution of overdensities in the
presence of Jeans smoothing.  Because the low-density IGM has a
dynamical time on the order of a Hubble time, it slowly reacts to any
photo-heating, and the filtering mass can be thought of some running
time-average of the Jeans mass.  Thus at some redshift, the Jeans and
filtering mass may differ substantially.  \citet{Thoul96} originally
showed photo-heating could totally suppress any star formation in
low-redshift dwarf galaxies with $V_c \lsim 35 \kms$ and somewhat
lower SFRs in galaxies up to 100 \kms.  At $z \gsim 3$, halos are less
susceptible to negative feedback from photo-heating because of their
higher average densities and thus cooling rates \citep{Dijkstra04}.
Closer inspection of collapsing halos in radiation hydrodynamics
simulations also shows high-redshift, low-mass halos can collapse and
is indeed regulated by the filtering mass.  Prior to reionization,
\citet{Gnedin00} showed that the filtering mass smoothly increases
from $\sim$\tento{6}~\Ms~before any star formation occurs to
\tento{9}\Ms~at $z = 6$.  \citet{Wise08b} calculated the filtering
mass in regions that were ionized by Population III stars and found
that it gradually increases the filtering mass to $\sim$$3 \times 10^7
\Ms$ at $z \sim 15$ around biased regions.

Recently, \citet{Mesigner08} studied how photo-heating from an
inhomogeneous UVB suppresses gas cooling and star formation.  They
found that at $z = 10$ an UVB intensity of $J_{21} \sim 0.1$ is
necessary to completely suppress gas collapse in $\mvir = 10^8 \Ms$
halos, where $J_{21}$ is in units of \tento{-21}~\emis.  In their
semi-numerical simulations at $z = 10$, they showed this intensity
occurs in a volume fraction of $<1\%$, suggesting that star formation
is widespread in these low-mass halos even in the advanced stages of
reionization.  They also illustrate how the collapsed gas fraction
steadily decreases with increasing UVB.  Even at $z = 7$ in a
\tento{8}\Ms~halo, approximately 40\% of the gas is retained when
compared to the no feedback case.  This gas mass loss will directly
affect star formation and perhaps the subsequent UV escape fraction.
Using the results of \citeauthor{Mesigner08} and the dependence of
\fescs~on gas fraction (see Fig. \ref{fig:fesc_cstar_trends}), our
results from cosmological halos can be adjusted to account for
external feedback from the UVB.  For example, at $z = 10$ with their
fiducial model, their semi-numerical simulations show that $J_{21}
\sim 0.01$ is most common, which decreases the gas fraction of
\tento{8}, \tento{8.5}, \tento{9}\Ms~halos to 0.2, 0.6, 0.85,
respectively, of the gas fraction without any UVB.  From
Fig. \ref{fig:fesc_cstar_trends}, we can see that \fescs~drops by
$\sim$50\% when the gas fraction is $\le$0.075 in idealized halos.  We
would expect a similar decrease in halos that lose gas from UVB
feedback, resulting in $\fescs \sim 0.02$ for halos in this mass
range.


\subsection{Prior Star Formation Episodes}

We assumed that the halos were unaffected by any prior stellar
feedback, but in this paper, we have also shown that low-mass halos
are susceptible to gas ``blow-out'' caused by radiative feedback.
This effect should be evident starting with Population III star
formation \citep[e.g.][]{Yoshida07, Wise08a} in halos with circular
velocities up to $V_c \sim 35 \kms$.  For example,
\citeauthor{Wise08a} found that gas fractions of dwarf galaxies with
$\mvir \sim 3 \times 10^7 \Ms$ at redshift 15 have been decreased to
$\le0.1$.  The cosmological halos presented here, which have $f_b \sim
0.13$, have $f_{\rm esc} \sim 0.5$ with a top-heavy IMF; however if
$f_b$ were lower from previous gas ``blow-out'', the trends shown in
Figure \ref{fig:trends} suggest that the escape fraction in these
halos should be even higher, but their values of
\fescs~(Fig. \ref{fig:fesc_cstar_trends}) will decrease because of
lower SFRs.  To further understand and quantify the consequences of
these neglected processes, semi-analytical reionization models,
semi-numerical or N-body + radiative transfer simulations are needed,
as discussed in \S\ref{sec:models}.  In future work, we plan to study
this in detail by following the hierarchical assembly of high-redshift
dwarf galaxies in cosmological simulations, including both metal-free
and metal-enriched star formation and feedback.

\section{Summary}
\label{sec:summary}

We presented results from an extensive suite of very high resolution
($0.1~$pc) AMR radiation hydrodynamics simulations that focus on the
UV escape fraction from isolated halos and cosmological halos.  The
latter cases were extracted from two large-scale cosmological
simulations.  These simulations accurately track the evolution of
\ion{H}{2} regions while including radiative and SNe feedback on the
surrounding medium.  The halo masses studied in this work span from $3
\times 10^6$ to $3 \times 10^9 \Ms$.  We used the isolated halos to
gauge if the escape fraction depends on the turbulent fractional
energy, spin parameter, and baryon mass fraction of the halo.  The
cosmological halos provide a good estimate of realistic escape
fractions of high-redshift dwarf galaxies.  We also investigated how a
top-heavy IMF and a normal IMF affects the escape fraction.  Our main
findings in this paper are

\medskip

1. Radiative feedback from massive stars, primarily arising from
D-type fronts, is most effective at ejecting gas from halos with
masses less than $10^{8.5} \Ms$ ($V_c = 35 \kms$).

2. Radiation preferentially escapes through channels with low column
densities, creating anisotropic \ion{H}{2} and a highly varying
$f_{\rm esc}$ along different lines of sight, agreeing with previous
work.

3. For a given halo mass, escape fractions in isolated halos have a
standard deviation of 0.2, arising from differing physical halo
parameters.  The largest effect comes from the baryon mass fraction,
where $f_{\rm esc}$ is lower in gas-rich halos with masses below
\tento{8} \Ms.  In gas-rich halos with masses larger than \tento{8}
\Ms, star formation is more intense, overcoming any large neutral
fraction it must ionize and resulting in higher values of $f_{\rm
  esc}$ than halos with smaller gas fractions.

4. High-redshift dwarf galaxies with $\mvir > 10^7 \Ms$, a top-heavy
IMF, and irregular morphology have $0.25 \le f_{\rm esc} \le 0.8$,
which we determined from our simulations of cosmological halos.  A
normal IMF decreases $f_{\rm esc}$ to 0.05--0.1 in halos with $\mvir <
10^{7.5} \Ms$ and $f_{\rm esc} \sim 0.4$ in more massive halos, which
have maximum SFRs spanning almost 2 orders of magnitude.

5. Escape fractions are dependent not only on the current SFR but on
the photo-heating and dispersion of gas, following feedback from
previous episodes of star formation.  Values of $f_{\rm esc}$ can vary up
to an order of magnitude in a few million years, i.e. the dynamical
time of a molecular cloud on which variations in SFRs can occur.

6. The mean of the product of star formation efficiency and ionizing
photon escape fraction, averaged over all atomic cooling ($\tvir \ge
8000~$K) galaxies, ranges from $0.02$ for a normal IMF to $0.03$ for a
top-heavy IMF.  Smaller, molecular cooling galaxies in minihalos are
not significant contributors to reionizing the universe primarily
because of a much lower star formation efficiency in minihalos than in
atomic cooling halos.

\medskip

The high escape fraction of UV radiation has important implications on
reionization, allowing a large amount of radiation from low-luminosity
dwarf galaxies to freely propagate into the IGM.  Although our
simulations miss the entire assembly of halos and their complete star
formation history, they are robust in following the dynamics of star
formation and feedback during the 100 Myr studied and suggest that
escape fraction in these galaxies are larger than previously assumed.

\acknowledgments

We thank an anonymous referee for useful suggestions.  J.~H.~W. thanks
Tom Abel for useful discussions.  We performed these calculations on
Orange at SLAC, Discover at NASA/GSFC, Abe and Cobalt at NCSA, and
Queenbee at LSU.  This research was supported by an appointment to the
NASA Postdoctoral Program at the Goddard Space Flight Center,
administered by Oak Ridge Associated Universities through a contract
with NASA.  We gratefully acknowledge financial support by grants
AST-0407176 and NNG06GI09G.  The computing time at NCSA was provided
by LRAC allocation TG-MCA04N012.

{}

\end{document}